\documentclass[10pt,twocolumn,aps,pra,showpacs,superscriptaddress,floatfix,longbibliography]{revtex4-1}

\usepackage[T1]{fontenc}
\usepackage{graphicx}
\usepackage[utf8]{inputenc}
\usepackage{amsmath,amsthm,amssymb,amsfonts,mathtools}
\usepackage{color}
\usepackage{psfrag}
\usepackage{epsfig}
\usepackage{bbm}
\usepackage{bm}
\usepackage[colorlinks=true,citecolor=blue,urlcolor=magenta]{hyperref}
\usepackage[normalem]{ulem}
\usepackage{epstopdf}
\usepackage{graphicx}
\usepackage{stackengine,xcolor}
\usepackage{comment}
\usepackage{hyperref}
\PassOptionsToPackage{linktocpage}{hyperref}
\usepackage[capitalize]{cleveref}
\usepackage[caption=false]{subfig}

\definecolor{nred}{rgb}{0.9,0.1,0.1}
\definecolor{nblack}{rgb}{0,0,0}
\definecolor{nblue}{rgb}{0.2,0.2,0.8}
\definecolor{ngreen}{rgb}{0.2,0.6,0.2}

\usepackage{etoolbox}
\usepackage{bbold} 

\DeclarePairedDelimiter{\ket}{\vert}{\rangle}
\DeclarePairedDelimiterX\braket[2]{\langle}{\rangle}%
  {#1\kern0.15ex\delimsize\vert\kern0.15ex\mathopen{}#2}

\DeclarePairedDelimiterX\ketbra[2]{\vert}{\vert}%
  {#1\kern0.15ex\delimsize\rangle\delimsize\langle\kern0.15ex\mathopen{}#2}

\DeclarePairedDelimiterX{\abs}[1]{\lvert}{\rvert}{%
  \ifblank{#1}{\,\cdot\,}{#1}
}   
\DeclarePairedDelimiterX\norm[1]\lVert\rVert{%
  \ifblank{#1}{\,\cdot\,}{#1}
}   
\DeclarePairedDelimiterX\mean[1]\langle\rangle{%
  \ifblank{#1}{\,\cdot\,}{#1}
}   

\newcommand{\id}{\mathsf{id}}
\newcommand{\A}{\mathrm{A}}
\newcommand{\B}{\mathrm{B}}
\newcommand{\AB}{\mathrm{AB}}
\newcommand{\HH}{\mathcal{H}} 
\newcommand{\HA}{\mathcal{H}_\A} 
\newcommand{\HB}{\mathcal{H}_\B} 
\newcommand{\DM}{\mathcal{S}} 
\newcommand{\T}{\intercal} 
\newcommand{\LL}{\mathcal{L}}
\newcommand{\argdot}{\,\cdot\,}
\newcommand{\CC}{\mathbb{C}}

\newcommand{\kbpsi}{\ketbra{\psi^+}{\psi^+}}

\newcommand{\sax}{\sigma_{a\vert x}}
\newcommand{\hsax}{\hat\sigma_{a\vert x}}
\newcommand{\rx}{\rho_x}
\newcommand{\rab}{{\rho_\AB}}
\newcommand{\nx}{n_x}
\newcommand{\ny}{n_y}
\newcommand{\na}{n_a}
\newcommand{\nb}{n_b}
\renewcommand{\Pr}{P}
\newcommand{\Fid}{\mathrm{F}}
\newcommand{\rf}{\mathrm{ref}}
\newcommand{\opt}{\mathrm{opt}}
\newcommand{\vx}{x}

\providecommand\given{}
\newcommand\SetSymbol[1][]{%
  \nonscript\:#1\vert
  \allowbreak
  \nonscript\:
  \mathopen{}}
\DeclarePairedDelimiterX\Set[1]\{\}{%
\renewcommand\given{\SetSymbol[\delimsize]}   
#1
}

\DeclareMathOperator{\tr}{tr}

\theoremstyle{definition}

\usepackage[nolist,withpage]{acronym}
\newacro{DI}{device-independent}
\newacro{CHSH}{Clauser-Horne-Shimony-Holt}
\newacro{SDP}{semidefinite programming}
\newacro{CPTP}{completely positive trace-preserving}
\newacro{ONB}{orthonormal basis}
\newacroplural{ONB}{orthonormal bases}
\newacro{POVM}{positive operator-valued measure}
\newacro{PSD}{positive semidefinite}
\newacro{RAC}{random access code}

\AtBeginDocument{%
    \newwrite\bibnotes
    \def\bibnotesext{Notes.bib}
    \immediate\openout\bibnotes=\jobname\bibnotesext
    \immediate\write\bibnotes{@CONTROL{REVTEX41Control}}
    \immediate\write\bibnotes{@CONTROL{%
    apsrev41Control,author="08",editor="1",pages="1",title="0",year="1"}}
     \if@filesw
     \immediate\write\@auxout{\string\citation{apsrev42Control}}%
    \fi
}%

\begin{document}

\title{Universal method for optimized robustness in self-testing of quantum resources}

\author{Shin-Liang Chen}
\thanks{shin.liang.chen.at.email.nchu.edu.tw}
\affiliation{Department of Physics, National Chung Hsing University, Taichung 402, Taiwan}
\affiliation{Physics Division, National Center for Theoretical Sciences, Taipei 10617, Taiwan}
\affiliation{Center for Quantum Frontiers of Research \& Technology (QFort), National Cheng Kung University, Tainan 701, Taiwan}

\author{Nikolai Miklin} 
\affiliation{Institute for Quantum-Inspired and Quantum Optimization, Hamburg University of Technology, Germany}
\affiliation{Institute for Applied Physics, Technical University of Darmstadt, Darmstadt, Germany}

\begin{abstract}
Self-testing is a phenomenon where the use of specific quantum states or measurements can be inferred solely from the correlations they generate.
We introduce a universal method for conducting robustness analysis in the self-testing of various quantum resources.
Unlike previous numerical approaches, which rely on selecting specific isometries, our method optimizes over equivalence transformations, thereby leading to tighter robustness bounds. 
This optimization employs the well-established technique of semidefinite programming relaxations for non-commuting polynomial optimization. 
Our method can be universally applied to diverse self-testing settings, including steerable assemblages in the Bell scenario, constellations of quantum states in the prepare-and-measure scenario, and entangled states in the steering scenario. 
We demonstrate the method's capability to surpass previously reported robustness bounds across a range of concrete examples.
\end{abstract}

\maketitle

Certifying properties of quantum systems is one of the central problems in quantum technologies~\cite{Eisert2020,Kliesch2021theory}.
Certification methods can be categorized according to the degree of trust placed in the experimental apparatus. 
This categorization defines a spectrum, ranging from scenarios in which all devices are fully characterized to those in which virtually no assumptions are made about the devices.
The latter case is known as the \emph{\ac{DI}}~\cite{Acin07} certification, which has gained significant attention over the past decade due to its application in quantum key distribution~\cite{Acin07,Pironio10,Nadlinger2022}. 
Several quantum properties, such as entanglement~\cite{Wiseman07}, steerability~\cite{Wiseman07}, measurement incompatibility~\cite{Wolf09}, randomness~\cite{Pironio10}, and dimension~\cite{Gallego10}, have been demonstrated to be certifiable and even quantifiable in a \ac{DI} manner~\cite{Moroder13,CBLC16,CBLC18}.

Remarkably~\cite{Summers87,Popescu92,Tsirelson93}, not only can the properties of quantum resources be certified in a \ac{DI} way, but even the resources themselves, which have become known as \emph{quantum self-testing}~\cite{Mayers04}. 
Seminal works have investigated self-testing of entangled quantum states~\cite{Mayers04,Yang14,Kaniewski16} and incompatible measurements~\cite{McKague12_0,Kaniewski2017} in the Bell test~\cite{Bell64}.
Later, the idea of self-testing was extended to other, so-called \emph{semi-\ac{DI}} scenarios, including quantum steering~\cite{Uola2020Steering} and the prepare-and-measure scenario~\cite{Pawowski2011}, where other quantum resources, such as quantum assemblages~\cite{Chen2021robustselftestingof}, generalized measurements~\cite{Miklin2020,Mohan2019,Tavakoli2020self,Anwer2020}, quantum gates~\cite{Noller2025classical,Noller2025sound}, and collections of quantum states~\cite{Miklin2021universalscheme,Tavakoli2020Semi,Navascues2023,SLChen2024} and measurements~\cite{Tavakoli18b,Farkas19} were shown to be certifiable (see Ref.~\cite{Supic2020selftestingof} for a review on self-testing). 

In simple terms, self-testing occurs when only a specific set of quantum resources can be used to explain the observed statistics or achieve the highest possible score in a given task. 
Importantly, all elements in this set are connected through some equivalence transformations, e.g., local isometries.
For instance, the maximum quantum violation of the \ac{CHSH} inequality~\cite{Clauser69} can be achieved only if the underlying quantum state is \emph{effectively} a maximally entangled two-qubit state~\cite{Popescu92}. 
The latter means that a pair of local isometries can be found that map the underlying state to a two-qubit singlet. 
Demonstrating the existence of such isometries is a key part of a self-testing proof; however, often more challenging is showing that the self-testing property is \emph{robust}~\cite{Kaniewski16}. 

Yang et al.~\cite{Yang14} proposed a general technique, which became known as the \emph{SWAP method}, based on the well-established hierarchy of \ac{SDP} relaxations~\cite{NPA,NPA2008,Pironio10b,Navascues15prl,Navascues15} that can be used to derive robust self-testing statements for quantum states and measurements in the Bell scenario. 
In simple terms, robust self-testing provides bounds on how close the implemented operations are to the target quantum resource when the observed statistic or figure of merit deviates from the ideal case.
Despite the success of the SWAP method, as pointed out by the authors themselves, the constructed SWAP transformation may not be optimal~\cite{Yang14}. 
In an accompanying work~\cite{Bancal15}, the authors proposed tweaking the SWAP method, which allows for some improvement in the robustness bounds.
However, to our knowledge, no systematic method for constructing an optimal equivalence transformation in robustness analysis has been proposed.

\begin{figure*}[t!]
\subfloat[\label{Fig_ST_assemblage}]{%
\includegraphics[width=4.2cm]{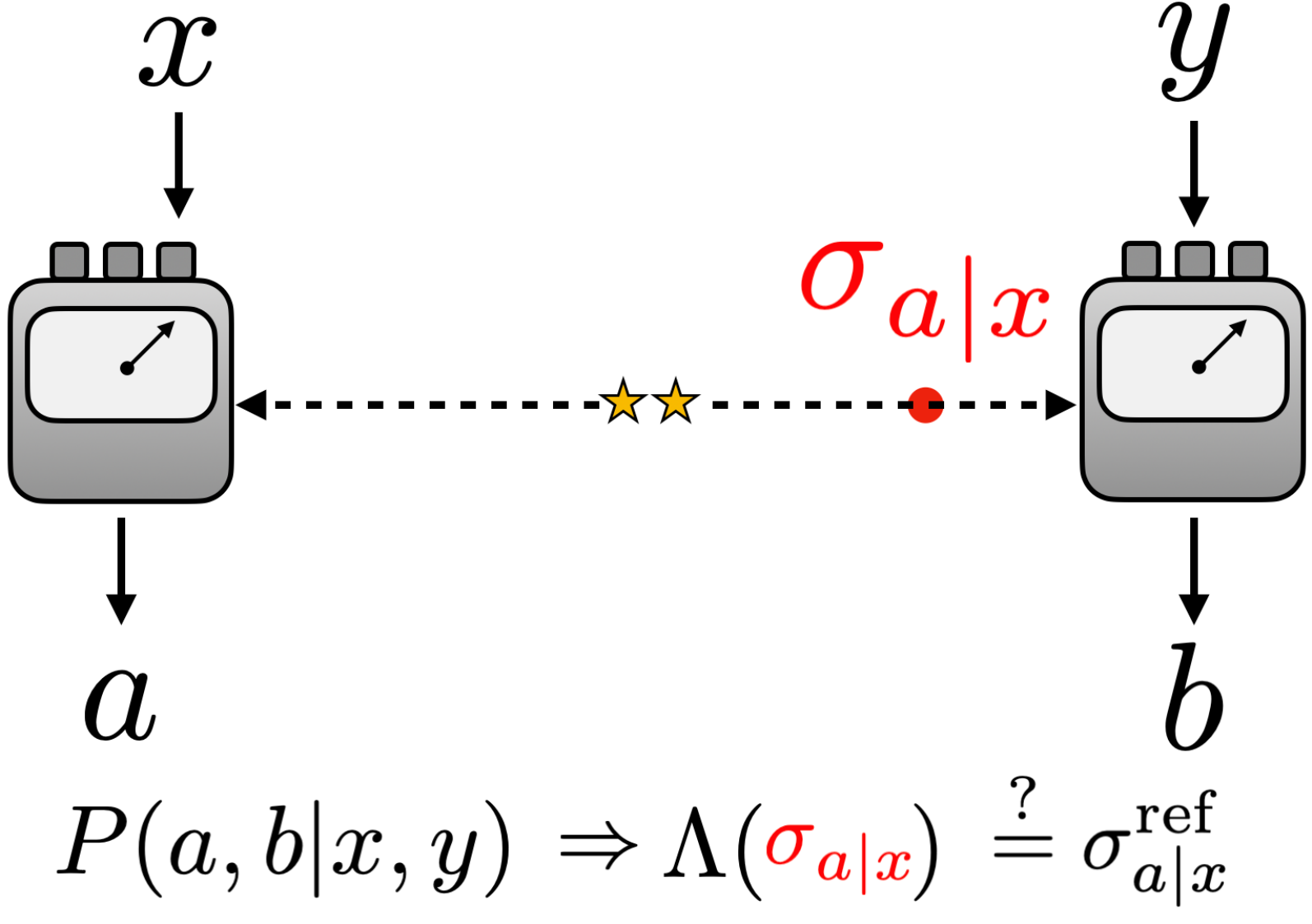}%
}\hspace{.1\linewidth}
\subfloat[\label{Fig_ST_PMstate}]{%
\includegraphics[width=4.2cm]{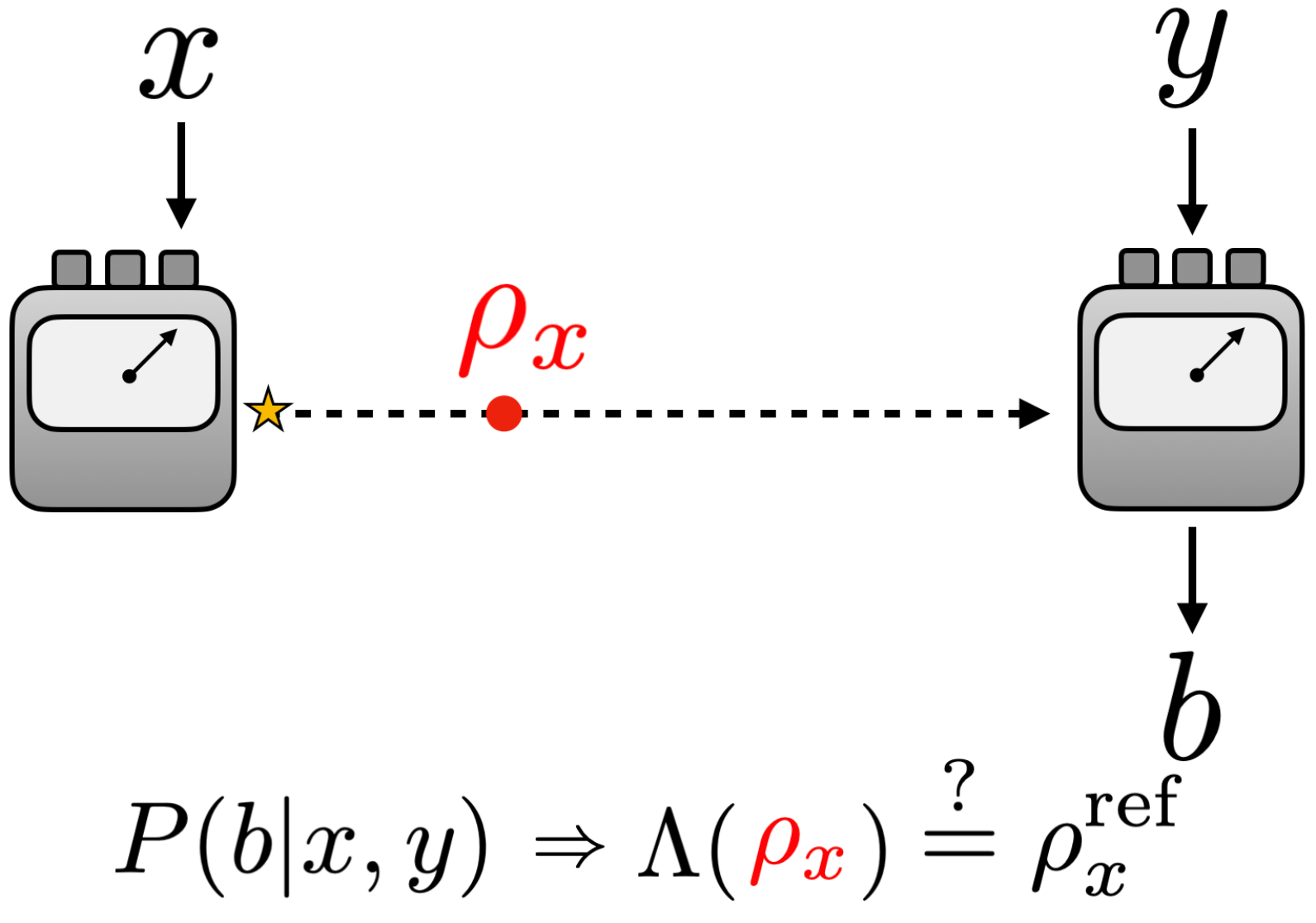}%
}\hspace{.1\linewidth}
\subfloat[\label{Fig_ST_1SDI}]{%
\includegraphics[width=4.2cm]{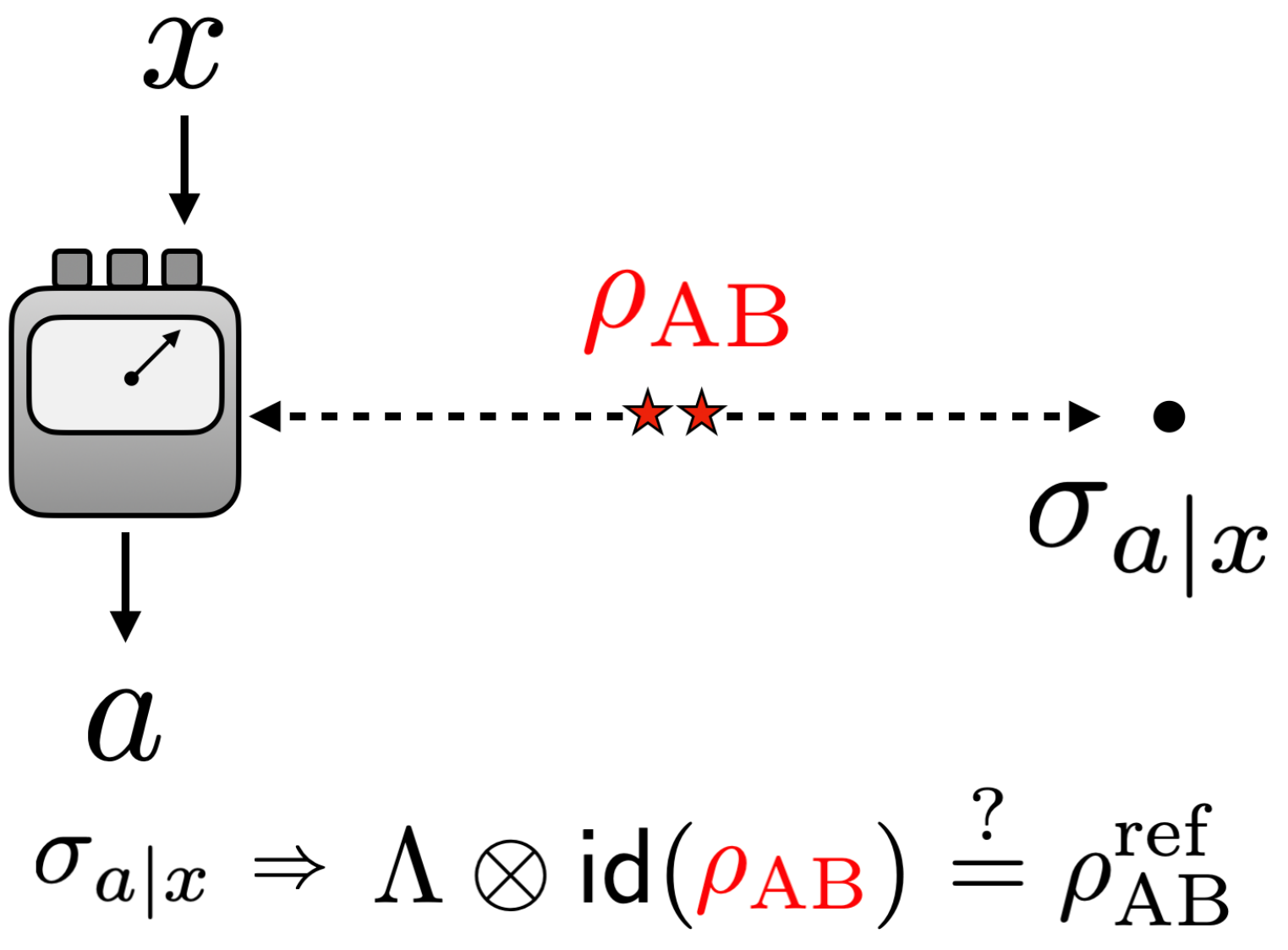}%
}
\caption{
The considered self-testing settings:
(a) steerable assemblages $\Set{\sax}_{a,x}$ in the Bell scenario, (b) constellations of states $\Set{\rx}_x$ in the prepare-and-measure scenario, (c) entangled states $\rab$ in the steering scenario. 
In all the considered two-party scenarios, $x\in [\nx]$, $a\in [\na]$ and $y\in [\ny]$, $b\in [\nb]$, are Alice's and Bob's inputs and outputs, respectively.
The central question in self-testing, is whether from the observed data $\Delta$, namely $\Pr(a,b\vert x,y)$ in (a), $\Pr(b\vert x,y)$ in (b), and $\sax$ in (c), it follows that there exists an equivalence transformation $\Lambda$ that maps the reference resource to the experimental one.
}
\label{Fig_overview}
\end{figure*}

In this work, we aim to fill this gap. 
To that end, we extend the framework of \ac{SDP} relaxations for non-commuting polynomial optimization~\cite{NPA,NPA2008,Pironio10b,Navascues15prl,Navascues15,Tavakoli2023Review}, previously applied to device-independent quantification of entanglement~\cite{Moroder13}, steerability~\cite{CBLC16,CBLC18}, measurement incompatibility~\cite{CMBC2021}, and characterization of temporal correlations~\cite{SLChen2024}.
For a range of self-testing settings, we show that optimization over equivalence transformations can be seamlessly incorporated into the hierarchy of \ac{SDP} relaxations for optimization over the underlying quantum resources.
More precisely, we consider self-testing of: i) steerable assemblages in the Bell scenario, ii) constellations of quantum states in the prepare-and-measure scenario, and iii) entangled states in the steering scenario.
For the quantum resources considered, the resulting robustness bounds are tighter than previously reported and, in some cases, optimal, being attained by explicit quantum realizations.

\emph{Preliminaries.---}
We use a shorthand notation $[n]\coloneqq \Set{0,1,\dots n-1}$ for a positive integer $n$.
In this work, we consider three bipartite scenarios as shown in \cref{Fig_overview}.
In each of the considered scenarios, $x\in [\nx]$, $a\in [\na]$, and $y\in [\ny]$, $b\in [\nb]$ are Alice's and Bob's inputs and outputs, respectively.
We denote the finite-dimensional Hilbert space associated with Alice and Bob's quantum system as $\HH$, with $\HH=\HA\otimes\HB$ if each party operates only on its part of the quantum system.
The sets of quantum states and linear operators on $\HH$ are denoted as $\DM(\HH)$ and $\LL(\HH)$, respectively.

In this paper, we focus on the self-testing of quantum states and state-like quantum resources (see \cref{Fig_overview}).
More precisely, we consider (a) steerable assemblages $\Set{\Pr(a\vert x)\hsax}$, with $\hsax\in\DM(\HB)$, and $\Pr(a\vert x)$ -- conditional probabilities, (b) constellations (sets) of quantum states $\Set{\rx}_{x\in[\nx]}$, with $\rx \in \DM(\HH)$, and (c) entangled states $\rab \in \DM(\HA\otimes\HB)$.
Let us, for the moment, denote a quantum resource as $\bm{\rho}$, having in mind either of the three cases.

The general goal of robust self-testing is to say how close the physical quantum resource $\bm{\rho}$, i.e., one used in the experiment, is to the reference one $\bm{\rho}^\rf$, given the observed data $\Delta$, e.g., a probability distribution, up to equivalence transformations, e.g, local isometries~\cite{Supic2020selftestingof}.
Following previous works on self-testing~\cite{Kaniewski16,Tavakoli18b,Chen2021robustselftestingof,SLChen2024}, as the class of equivalence transformations, we consider \ac{CPTP} maps applied to the physical resource.
As the metric of ``closeness'' for $\bm{\rho}$ and $\bm{\rho}^\rf$, we take the (averaged) fidelity, which in the cases we consider takes the form
\begin{equation}\label{Eq_Def_Fid} 
    \Fid \coloneqq \sum_{i\in I}c_i\tr\left[\rho_i^\rf\Lambda(\rho_i)\right],
\end{equation}
where $I$ is a finite set, $c_i\geq 0$, $\rho^\rf_i\in \LL(\HH')$, $\rho_i\in \LL(\HH)$ are ``parts'' of the reference and physical resources, respectively, and $\Lambda: \LL(\HH)\to\LL(\HH')$ is a \ac{CPTP} map.
Note that $\HH$ and $\HH'$ may have different dimensions.

Given the considered type of quantum resources and the chosen metric~\eqref{Eq_Def_Fid} together with the class of equivalence transformations, the task of establishing a robust self-testing result reduces to the following optimization problem (see also~\cite{Kaniewski16,Tavakoli18b}):
\begin{equation}
\begin{aligned}
\Fid_\opt \coloneqq \min_{\bm{\rho}} \max_{\Lambda}\quad & \sum_{i\in I} c_i\tr\left[\rho_i^\rf\Lambda(\rho_i)\right]\\
\text{subject to}\quad & \rho_i\succeq 0,\quad \forall i\in I,\\
&\bm{\rho}~\text{compatible with}~\Delta,\\
& \Lambda~\text{CPTP}.
\end{aligned}
\label{Eq_Def_Fopt}
\end{equation}
In words, in \cref{Eq_Def_Fopt}, we look for the worst resource $\bm{\rho}$ compatible with the observed data $\Delta$, up to equivalence transformations $\Lambda$.
Note that a restricted type of \ac{CPTP} maps, e.g.~unitary transformations, is sometimes considered~\cite{Miklin2021universalscheme,Noller2025classical,Navascues2023}, and an additional complex conjugation applied to $\rho_i^\rf$ with respect to a chosen basis may be required to capture all types of equivalence transformations~\cite{wigner1931gruppentheorie}.

Currently, no algorithm is known that guarantees finding the optimal $\Fid_\opt$ in \cref{Eq_Def_Fopt}.
It is also far from obvious how to apply the hierarchy of \ac{SDP} relaxations~\cite{NPA,NPA2008,Pironio10b,Navascues15prl,Navascues15} in this setting. The difficulty for the latter is twofold. First, the problem in \cref{Eq_Def_Fopt} is a min–max optimization. 
Second, the objective function is not a linear combination of the moments of the operators over which the optimization is performed.
The proposal of the SWAP method~\cite{Yang14,Bancal15} is to choose a particular map $\Lambda$, and express its input space part in terms of the operators of optimization, thus fitting within the framework of the \ac{SDP} relaxations~\cite{NPA,NPA2008,Pironio10b,Navascues15prl,Navascues15}, although, at the cost of further relaxing the problem.

\emph{Universal method for optimized robustness.--} 
We present an alternative strategy to address the aforementioned challenges associated with the optimization problem \eqref{Eq_Def_Fopt} for a range of self-testing scenarios.
Whenever the observed data $\Delta$ admits a quantum realization, the \ac{SDP} in \cref{Eq_Def_Fopt} with respect to $\Lambda$ is feasible (and in practice strictly feasible).
Moreover, the objective function is bounded from below and above by $0$ and $\sum_{i\in I}c_i$, respectively.
Thus, strong duality of the optimization problem with respect to $\Lambda$ holds~\cite{Skrzypczyk2023SDP}, and we can take its dual \ac{SDP}:
\begin{equation}
\begin{aligned}
\Fid_\opt = \min_{\bm{\rho},Y} \quad & \tr(Y)\\
\text{subject to}\quad & Y\otimes\openone\succeq\sum_{i\in I} c_i\rho_i\otimes (\rho_i^\rf)^\T,\\
& \rho_i\succeq 0,\quad \forall i\in I,\\
&\bm{\rho}~\text{compatible with}~\Delta,
\end{aligned}
\label{Eq_Fopt_dual}
\end{equation}
where $Y\in \LL(\HH)$ (see \cref{SecApp_Eq_Fopt_dual} for details).
One direct advantage of taking the dual of \eqref{Eq_Def_Fopt} is that we no longer have to deal with the min-max optimization. 

Next, we show how to fit \eqref{Eq_Fopt_dual} within the framework of \ac{SDP} relaxations for quantum correlations~\cite{NPA,NPA2008,Pironio10b,Navascues15prl,Navascues15,Tavakoli2023Review} for various types of quantum resources~\cite{Moroder13,Pusey13,Kogias15,CMBC2021,SLChen2024}. 
The main procedure behind this method is constructing the so-called \emph{moment matrix} by applying a completely positive map $\Gamma$ on the underlying quantum resource, more precisely, each $\rho_i$, \emph{and} the dual variable $Y$ in \cref{Eq_Fopt_dual}.
Let $d\coloneqq \dim(\HH)$, and $\Set{S_j\in \LL(\HH) \given j\in [m]}$ be a set of $m$ linear operators on $\HH$.
Then, we define $\Gamma$ as 
\begin{equation}\label{Eq_Def_Gamma}
    \Gamma(\argdot)\coloneqq \sum_{l\in [d]}K_l(\argdot)K_l^\dagger,
\end{equation}
with $K_l\coloneqq \sum_{j\in [m]}\ketbra{j}{l}S_j$, where $\Set{\ket{l}}_{l\in [d]}$ and $\Set{\ket{j}}_{j\in [m]}$ are \acp{ONB} in $\HH$, and $\CC^m$, respectively (see also~\cite{Moroder13,Pusey13,CBLC16,CMBC2021,SLChen2024}).
Direct calculation shows that $\Gamma(\rho_i) = \sum_{j,k\in [m]}\ketbra{j}{k}\tr\left(S_j^\dag S_k \rho_i\right)$, for $i\in I$.
If we choose the set of operators $\Set{S_j}_{j\in[m]}$ to contain physical quantum states and \ac{POVM} effects, then the elements of $\Gamma(\rho_i)$ are going to be their moments.
Since $\Gamma$ is a completely positive map, it preserves the positivity of operators, and therefore, $\Gamma(\rho_i)\succeq 0$, for all $i\in I$.
Crucially, it also preserves the positive semidefinite constraints, such as the first constraint in \cref{Eq_Fopt_dual}, even if it is applied only to one tensor factor.
Finally, some elements of the constructed moment matrices correspond to the observed data $\Delta$, which allows us to accommodate the second constraint in \cref{Eq_Fopt_dual}.

The construction described above leads to the following relaxation of the optimization problem~\eqref{Eq_Fopt_dual}:
\begin{equation}
\begin{aligned}
\Fid_{\rm opt} \geq \min_{\mathbf{u}} \quad & \Gamma(Y)_{\openone}\\
\text{subject to}\quad & \Gamma(Y)\otimes\openone\succeq\sum_{i\in I} c_i\Gamma(\rho_i)\otimes (\rho_i^\rf)^\T,\\
&\Gamma(\rho_i)\succeq 0,\quad\forall i\in I,\\
& \Set{\Gamma(\rho_i)}_{i\in I} ~\text{compatible with}~\Delta,
\end{aligned}
\label{Eq_Fopt_final}
\end{equation}
where $\Gamma(Y)_{\openone}$ is the element of $\Gamma(Y)$ corresponding to the moment $\tr(Y)$, and by $\mathbf{u}$ we denote all the elements of $\Gamma(\rho_i)$ and $\Gamma(Y)$, which are not fixed by the observed data $\Delta$ and, thus, are variables of optimization.
The solution to the problem~\eqref{Eq_Fopt_final} provides a lower bound on $\Fid_\opt$, because any feasible solution of the problem~\eqref{Eq_Fopt_dual} is also feasible for the problem in \cref{Eq_Fopt_final}.
The main advantage of considering the relaxation~\cref{Eq_Fopt_final} is that it is an \ac{SDP} and thus can be efficiently solved on a personal computer with available packages~\cite{Lofberg2004,SDPT3}, with a guarantee of convergence~\cite{BoydBook}.
Note that there is also a possibility of restricting the dimension of the Hilbert space $\HH$ in \cref{Eq_Fopt_final} via construction of Refs.~\cite{Navascues15prl,Navascues15}.
In the next sections, we demonstrate that the proposed approach can be used to obtain robust self-testing results for a range of quantum resources. 

\emph{Self-testing of steerable assemblages in the Bell scenario.---}
In the Bell scenario (see \cref{Fig_ST_assemblage}), Alice and Bob share a bipartite state $\rab\in \DM(\HA\otimes\HB)$, and Alice performs a measurement, obtaining the outcome $a\in [\na]$ with probability $\Pr(a\vert x)$, with $x\in [\nx]$ specifying the choice of her measurement. 
Let us denote the resulting state of Bob's subsystem as $\hsax\in \DM(\HB)$.
The set of sub-normalized states $\Set{\sax}_{a\in[\na],x\in[\nx]}$, where $\sax\coloneqq \Pr(a\vert x)\hsax$, is known as the \emph{assemblage}~\cite{Uola2020Steering}.
An assemblage is said to be steerable if it does not admit a local-hidden-state model~\cite{Wiseman07}: $\sax=\sum_\lambda \Pr(\lambda)\Pr(a\vert x,\lambda)\sigma_\lambda$, where $\sigma_\lambda\in \DM(\HB)$, in which case it provides a resource and could, in principle, be self-tested.

Let Bob's measurement be described by \acp{POVM} $\Set{E_{b\vert y}}_{b\in[\nb]}$, with $y\in[\ny]$.
Then, the observed data $\Delta$ in the Bell scenario satisfies $\Pr(a,b\vert x,y) = \tr(\sax E_{b\vert y})$, with $\Pr(a\vert x)$ being its marginal.
Let us assume $\Pr(x)=\frac{1}{\nx}$, and let $\Set{\sax^\rf}_{a\in[\na],x\in[\nx]}$ be the reference assemblage with $\Pr^\rf(a\vert x)\coloneqq \tr(\sax^\rf)$.
The task of establishing robust self-testing of steerable assemblages in the Bell scenario can then be cast in the form~\eqref{Eq_Def_Fopt} by taking
\begin{equation}
\begin{aligned}
I&\rightarrow [\na]\times[\nx],& \Delta&\rightarrow \Pr(a,b\vert x,y),\\
c_i&\rightarrow \frac{1}{\nx}\frac{1}{\sqrt{\Pr(a\vert x)\Pr^\rf(a\vert x)}},& 
\rho_i&\rightarrow\sax,
\label{Eq_subst_1}
\end{aligned}
\end{equation}
(see \cref{SecApp_SDP_ST_assemblage} for details, and also \cite{Chen2021robustselftestingof} for previous analysis).
To construct the moment matrix, we take the operators $S_j$ to be the \ac{POVM} effects $E_{b\vert y}$ and their products.

\begin{figure}
\includegraphics[width=0.95\linewidth]{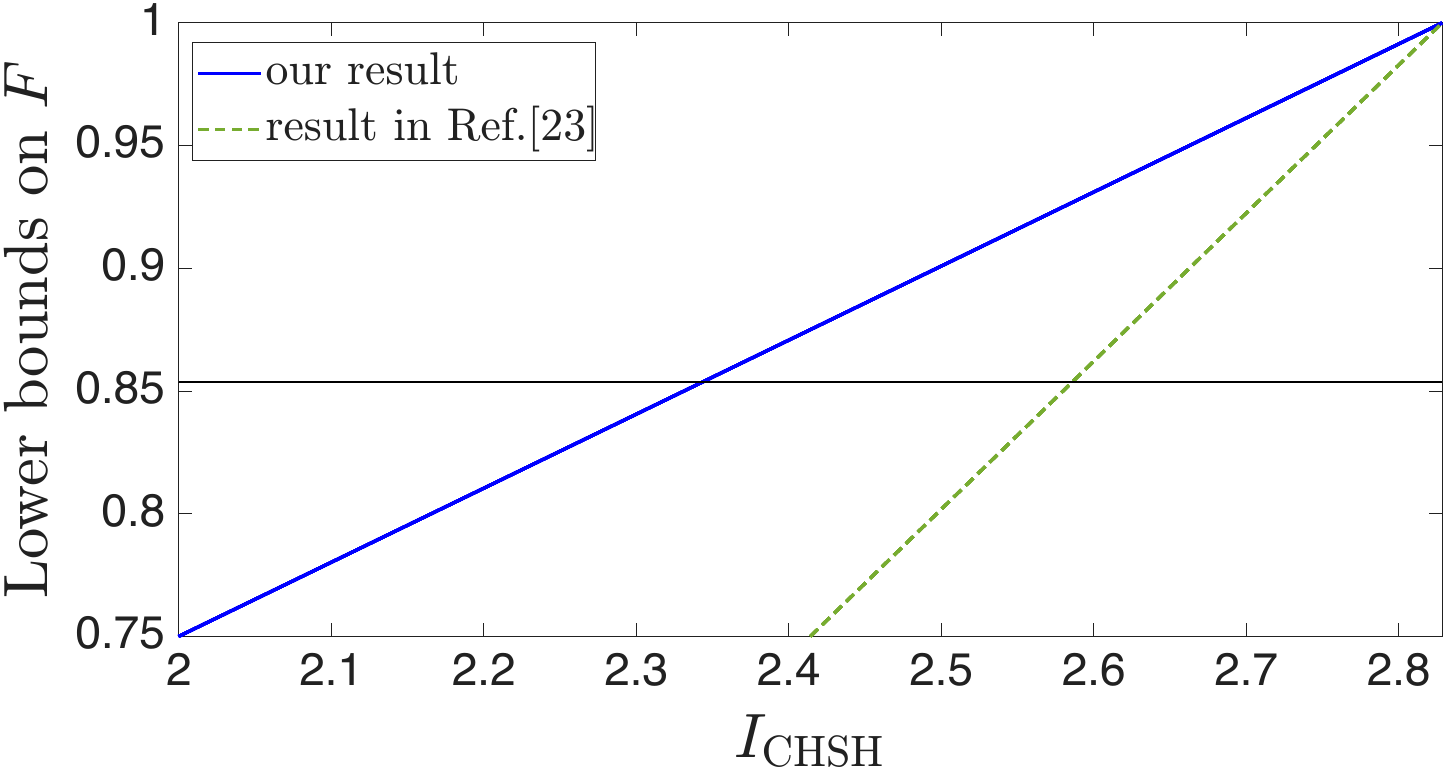}
\caption{Robust self-testing of a steerable assemblage in the \ac{CHSH} scenario. In this example, we assume $\Pr(a\vert x)=1/2$. The level of hierarchy we implement is $\ell=3$. The horizontal line with the value of $0.8536$ represents the classical fidelity.
} 
\label{Fig_min_F_steering_CHSH}
\end{figure}

As an example, we consider the following reference assemblage: $\{\frac{1}{2}\ketbra{0}{0}, \frac{1}{2}\ketbra{1}{1}, \frac{1}{2}\ketbra{+}{+}, \frac{1}{2}\ketbra{-}{-}\}$, i.e., the associated states $\hsax^\rf$ are symmetrically distributed on the $X$-$Z$ plane of the Bloch sphere with uniform probability distribution. 
In the Bell scenario~\cite{Bell64}, it is common to take as the observed data $\Delta$ the value of a Bell inequality expression~\cite{Supic2020selftestingof}, e.g.~\ac{CHSH} inequality
$I_{\rm CHSH}\coloneqq \langle A_0 B_0\rangle + \langle A_0 B_1\rangle + \langle A_1 B_0\rangle - \langle A_1 B_1\rangle$,
where $\langle A_x B_y\rangle\coloneqq \sum_{a,b\in\Set{0,1}} (-1)^{a+b}\Pr(a,b\vert x,y)$, with $2$ being it local-hidden-variable bound~\cite{Clauser69}. 
In \cref{Fig_min_F_steering_CHSH} we plot the output of the \ac{SDP}~\eqref{Eq_Fopt_final} versus the value of $I_{\rm CHSH}$, which, as one can see, gives a significant improvement over previously reported lower bounds~\cite{Chen2021robustselftestingof}.
Moreover, the bound we obtain is, in fact, tight (see \cref{SecApp_SDP_ST_assemblage} for details).

\emph{Self-testing of constellations of quantum states in the prepare-and-measure scenario.---}
Next, we consider self-testing of sets of quantum states in the prepare-and-measure scenario (see \cref{Fig_ST_PMstate}).
Here, Alice prepares a quantum system in a state $\rx\in\DM(\HH)$, specified by $x\in [\nx]$, and sends it to Bob, who measures it, with the measurement for the choice $y\in [\ny]$ described by a \ac{POVM} $\Set{E_{b\vert y}}_{b\in [\nb]}$.
The set $\Set{\rx}_{x\in[\nx]}$ may be a quantum resource for some task such as \ac{RAC}~\cite{Ambainis1999}, and may be self-tested (under additional assumptions) by the observed data $\Pr(b\vert \vx,y)=\tr(\rho_{\vx}E_{b\vert y})$~\cite{Tavakoli18b,Farkas19,Miklin2021universalscheme}.
We can cast the problem of robust self-testing of the preparation states in the prepare-and-measure scenario in the form~\eqref{Eq_Def_Fopt} by setting
\begin{equation}
I\rightarrow [\nx],\quad \Delta\rightarrow \Pr(b\vert \vx,y),\quad 
c_i\rightarrow \frac{1}{[\nx]},\quad \rho_i\rightarrow\rx,
\label{Eq_subst_2}
\end{equation}
(see \cref{SecApp_SDP_PM} for details, and also Ref.~\cite{Liang19}).
For the map~\eqref{Eq_Def_Gamma}, we choose the operators $S_j$ to be the \ac{POVM} effects of Bob $E_{b\vert y}$ as well as their arbitrary products.

\begin{figure}
\includegraphics[width=0.95\linewidth]{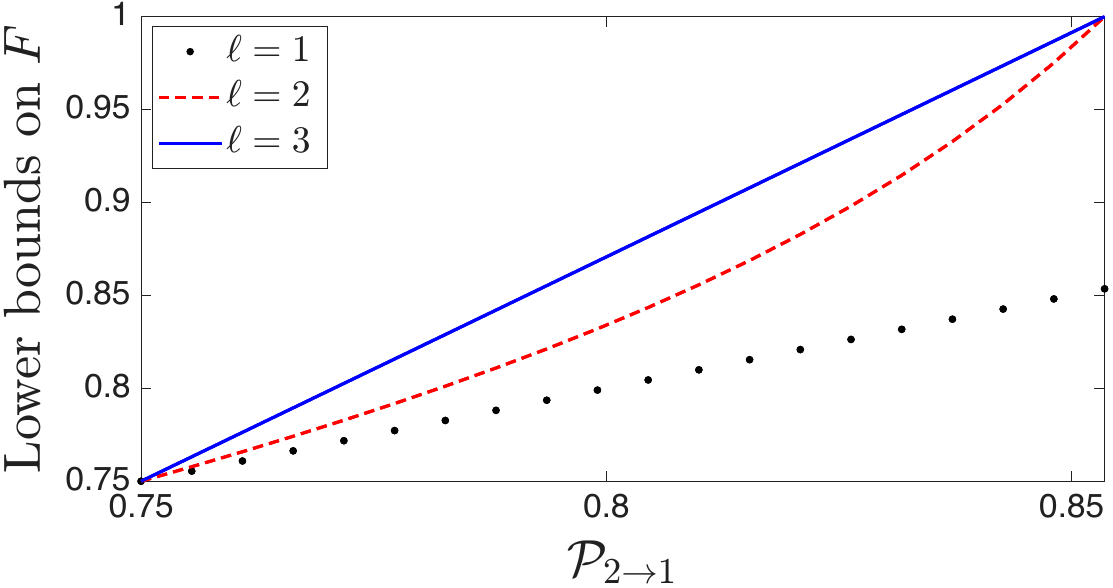}
\caption{Robust self-testing of a constellation of states in $2\to 1$ \ac{RAC} with shared randomness. $l\in\{1,2,3\}$ stands for the level of the \ac{SDP} hierarchy, defined similarly as in Refs.~\cite{Navascues15prl,Navascues15}.
}
\label{Fig_min_F_QRAC2to1}
\end{figure}

As a concrete example, we consider self-testing of preparation states in $2\to 1$ \ac{RAC} with shared randomness~\cite{ambainis2009quantumrandomaccesscodes} under the assumption of $\dim(\HH)=2$.
Here, the input of Alice is a $2$-bit string $\vx=(x_0,x_1)$, with $x_i\in\{0,1\}$, which is equivalent to having $\nx=4$.
Instead of the entire distribution $\Pr(b\vert \vx,y)$, we take the average success probability $\mathcal{P}_{2\to 1}\coloneqq\frac{1}{8}\sum_{\vx\in\Set{0,1}^2,y\in \Set{0,1}} \Pr(b\vert \vx,y)\delta_{b,x_y}$ as the observed data.
The maximal average successful probability $\mathcal{P}_{2\to 1}=\frac{1}{2}(1+1/{\sqrt{2}})$ can be achieved by the constellation of states with state vectors $\{\ket{0}, \ket{1}, \ket{+}, \ket{-}\}$~\cite{Ambainis1999}.
In \cref{Fig_min_F_QRAC2to1} we plot the output of \eqref{Eq_Fopt_final} versus the value of $\mathcal{P}_{2\to 1}$ for various levels of the hierarchy of \ac{SDP} relaxations~\cite{Navascues15prl,Navascues15}, where we add additional constraints to enforce the dimension of the quantum system to be $2$.
The resulting lower bound matches previously reported results in the literature (see e.g. Ref.~\cite{Tavakoli18b}), which are also known to be tight. 

\emph{Self-testing of entangled states in the steering scenario.---}
Finally, we show that our framework can also be applied to self-testing entangled states when only one of the measurement devices is treated as a black box, i.e., in the so-called steering or one-sided device-independent scenario (see \cref{Fig_ST_1SDI}).
In this scenario, the parties share an entangled state $\rab\in \DM(\HA\otimes\HB)$, and let this time $\Set{E_{a\vert x}}_{a\in [\na]}$ be the \ac{POVM} that describes Alice's measurement for the setting $x\in [\nx]$.
Here, the observed data $\Delta$ is the assemblage $\Set{\sax}_{a\in [\na],x\in[\nx]}$, connected to $\rab$ via $\sax = \tr_{\A}(E_{a\vert x}\otimes\openone \rab)$.
If $\sax$ is steerable, then the entangled state $\rab$ is a resource $\rab$, which can be self-tested~\cite{Supic2016}. 
To fit this self-testing setting into the form~\eqref{Eq_Def_Fopt}, we set
\begin{equation}
    I\to \Set{0},\quad \Delta\to\sax,\quad c_i\to 1,\quad \rho_i\to\rab,
\end{equation}
and, importantly, we restrict $\Lambda$ to act only on Alice's subsystem.
For the \ac{SDP} relaxation in \cref{Eq_Fopt_final}, we choose the operators $S_j$ to be the \ac{POVM} effects of Alice's measurements as well as their products.

\begin{figure}
\includegraphics[width=0.95\linewidth]{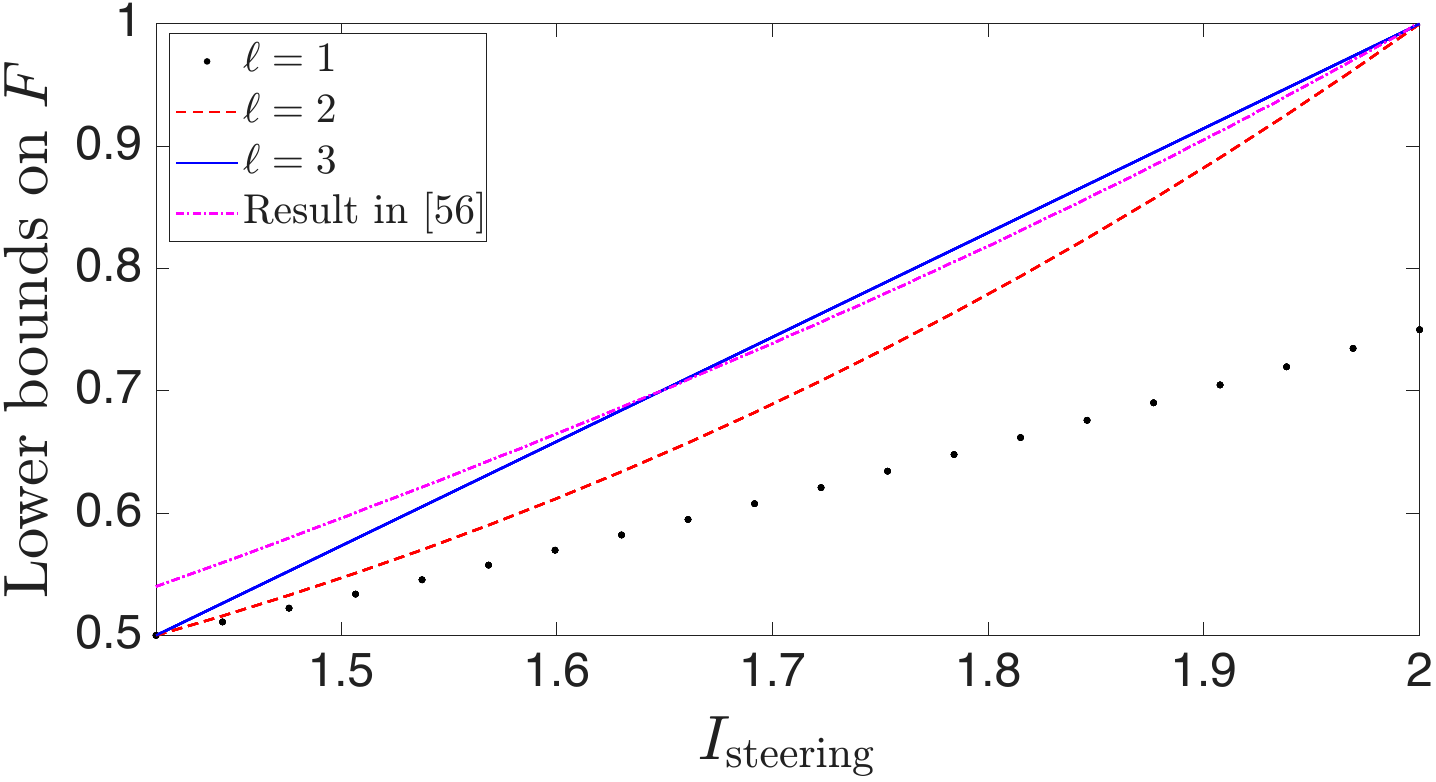}
\caption{Robust self-testing of a two-qubit entangled state in the steering scenario. $l\in\{1,2,3\}$ stands for the level of the \ac{SDP} hierarchy, defined similarly as in Refs.~\cite{NPA,NPA2008}.   
}
\label{Fig_minF_1SDI}
\end{figure}

As an example, we consider the steering scenario with binary measurement settings and outcomes, and in \cref{Fig_minF_1SDI} we plot the output of the \ac{SDP}~\eqref{Eq_Fopt_final} against the value of the steering inequality~\cite{Pusey13} (see also \cite{Supic2016}), which has the form $I_{\rm steering} \coloneqq \tr\left(Z(\sigma_{0\vert 0}-\sigma_{1\vert 0}) + X(\sigma_{0\vert 1}-\sigma_{1\vert 1})\right)$, where $Z$ and $X$ are the Pauli matrices.
The local-hidden-state bound of $I_{\rm steering}$ is $\sqrt{2}$, and the quantum bound is $2$.
As \cref{Fig_minF_1SDI} demonstrates, the lower bound on $\Fid_\opt$ we obtain improves on the previously reported values in Ref.~\cite{Supic2016} for $I_{\rm steering}\gtrsim 1.65$.

\emph{Discussion.---} In this work, we develop a method for robust self-testing of various quantum state-like resources. In contrast to previous approaches, our method does not require choosing a particular equivalence transformation, e.g., SWAP isometry, but allows one to optimize over them. 
We demonstrate our approach on three types of self-testing settings, and in the concrete examples we consider, the obtained robustness bounds show improvement over the previously reported results that, in some cases, are optimal.

The proposed approach to robust self-testing opens several promising avenues for future research. A natural first direction is to investigate how much improvement in terms of robustness can be achieved for examples beyond the simplest cases considered here. Moreover, the general idea of incorporating optimization over equivalence transformations into the \ac{SDP} relaxation could be extended to other quantum resources, such as measurements, as well as to different self-testing settings. We believe, however, that pursuing these extensions warrants a separate and dedicated study. Most notably, the self-testing of entangled states in the Bell scenario does not appear to fit directly into the form~\eqref{Eq_Def_Fopt}, as it requires enforcing the locality of two isometries.

\begin{acknowledgments}
S.-L.~C. acknowledges the support of National Science and Technology Council (NSTC) Taiwan (Grant No. NSTC 111-2112-M-005-007-MY4) and National Center for Theoretical Sciences Taiwan (Grant No.~NSTC 113-2124-M-002-003).
This research was funded by Fujitsu Germany GmbH as part of the endowed professorship ``Quantum Inspired and Quantum Optimization''.
\end{acknowledgments}

\onecolumngrid
\section*{APPENDIX}
\appendix
Here, we provide technical details that support the key statements from the main text.
In \cref{SecApp_Eq_Fopt_dual}, we provide details on the derivation of the dual optimization problem in \cref{Eq_Fopt_dual}.
In \cref{SecApp_SDP_ST_assemblage}, \cref{SecApp_SDP_PM}, and \cref{SecApp_SDP_ST_1SDI}, we provide details on the three types of self-testing settings that we consider in this paper (see \cref{Fig_overview}).

\section{Derivation of the dual problem in \cref{Eq_Fopt_dual}}
\label{SecApp_Eq_Fopt_dual}
Let us re-state the problem in \cref{Eq_Def_Fopt}:
\begin{equation}
\begin{aligned}
\text{given}\quad & \Set{c_i}_{i\in I}, \bm{\rho}^\rf, \Delta\\
\Fid_\opt = \min_{\bm{\rho}} \max_{\Lambda}\quad & \sum_{i\in I} c_i\tr(\rho_i^\rf\Lambda(\rho_i))\\
\text{subject to}\quad & \rho_i\succeq 0,\quad \forall i\in I,\\
&\bm{\rho}~\text{compatible with}~\Delta,\\
& \Lambda~\text{CPTP}.
\end{aligned}
\label{EqApp_Def_Fopt}
\end{equation}
Let the Choi operator~\cite{Jamiokowski74,Choi75} of the \ac{CPTP} map $\Lambda: \LL(\HH)\to \LL(\HH')$ be $\Omega\coloneq  (\id\otimes\Lambda)(\ketbra{\phi^+}{\phi^+})$, where $\mathsf{id}$ is the identity map and $\ket{\phi^+} \coloneqq \sum_{j\in [d]}\ket{j}\otimes\ket{j}\in \HH''\otimes\HH$, where $d\coloneqq \dim(\HH)=\dim(\HH'')$.
The condition of $\Lambda$ being \ac{CPTP} is equivalent to $\Omega\succeq 0$ and $\tr_{\HH}(\Omega)=\openone$.
It is easy to see that $\tr\left(\rho_i^\rf\Lambda(\rho_i)\right)=\tr\left(\rho^\T_i\otimes\rho_i^\rf\Omega\right)=\tr\left(\rho_i\otimes(\rho_i^\rf)^\T\Omega^\T\right)$.
Let $\bm{\rho}^\ast$ be a feasible point of \cref{EqApp_Def_Fopt}.
Then, the optimization problem with respect to $\Lambda$ in \cref{EqApp_Def_Fopt} is equivalent to the following
\begin{equation}
\begin{aligned}
\text{given}\quad & \Set{c_i}_{i\in I}, \bm{\rho}^\rf, \bm{\rho}^\ast\\
\max_\Omega\quad &\sum_{i\in I}c_i\tr\left(\rho_i\otimes(\rho_i^\rf)^\T\Omega\right)\\
\text{subject to}\quad & \Omega\succeq 0,\quad \tr_{\HH}(\Omega) = \openone,
\end{aligned}
\label{EqApp_max_over_Omega}
\end{equation}
where we changed the optimization variable from $\Omega$ to $\Omega^\T$.

To derive the dual problem, we start by writing the Lagrangian~\cite{BoydBook} (see also Ref.~\cite{Skrzypczyk2023SDP}):
\begin{equation}
\begin{aligned}
\mathcal{L} = \sum_{i\in I}c_i \tr\left(\rho^\ast_i\otimes(\rho_i^\rf)^\T\Omega\right) +
\tr\left[Y(\openone-\tr_\HH\Omega)\right] + \tr(Z\Omega)
= \tr\left[ \Omega \left(  \sum_{i\in I} c_i \rho^\ast_i\otimes(\rho_i^\rf)^{\T} -  Y\otimes\openone +Z \right)  \right]  + \tr Y,
\end{aligned}
\end{equation}
where $Y\in \LL(\HH'')$ and $Z\in \LL(\HH''\otimes\HH)$ are the dual variables, with $Z\succeq 0$.
One can see that $\mathcal{L}\geq \sum_{i\in I} c_i \tr\left(\rho_i\otimes(\rho_i^\rf)^\T \Omega\right)$ if $Z\succeq 0$, i.e., in other words, $\mathcal{L}$ upper-bounds the objective function of the primal problem. The dual problem corresponds then to the minimal upper bound, which amounts to
\begin{equation}
\begin{aligned}
\min_{Y,Z}\quad &\tr(Y)\\
\text{subject to}\quad & \sum_{i\in I} c_i \rho^\ast_i\otimes(\rho_i^\rf)^\T -  Y\otimes\openone +Z = 0,\\
& Z\succeq 0.
\end{aligned}
\end{equation}
Resolving the equality constraint with respect to $Z$ yields
\begin{equation}
\begin{aligned}
\min_{Y}\quad &\tr(Y)\\
\text{subject to}\quad & Y\otimes\openone \succeq \sum_{i\in I} c_i \rho^\ast_i\otimes(\rho_i^\rf)^\T.
\end{aligned}
\label{EqApp_Dual_of_Max}
\end{equation}
Including the minimization over $\bm{\rho}$ gives the optimization problem in \cref{Eq_Fopt_dual} in the main text.

\section{Details of robust self-testing of steerable assemblages in the Bell scenario}
\label{SecApp_SDP_ST_assemblage}
We start by writing the definition of fidelity between two assemblages, $\{\Pr(a\vert x)\hsax\}_{a\in[\na],x\in[\nx]}$ and $\{\Pr^\rf(a\vert x)\hsax^\rf\}_{a\in[\na],x\in[\nx]}$ (see e.g.,~ Ref.~\cite{Chen2021robustselftestingof}), up to the equivalence \ac{CPTP} map $\Lambda$:
\begin{equation}
\Fid = \frac{1}{\nx} \sum_{a\in [\na],x\in[\nx]}\frac{1}{\sqrt{\Pr(a\vert x) \Pr^\rf(a\vert x)}} \tr\left[\Lambda(\sax)\sax^\rf\right],
\end{equation}
where we assumed that $\hsax^\rf$ are pure states for all $a\in[\na],x\in[\nx]$.
Compared with the form of fidelity between two resources in \cref{Eq_Def_Fid}, we obtain the form of the index set $I$, the coefficients $c_i$, and states $\rho_i$ in \cref{Eq_subst_1}.
With this, the optimal fidelity in \cref{Eq_Fopt_dual} takes the following form
\begin{equation}
\begin{aligned}
\Fid_\opt = \min_{\sax,Y,E_{b\vert y}} \quad & \tr(Y)\\
\text{subject to}\quad & Y\otimes\openone\succeq \frac{1}{\nx} \sum_{a\in[\na],x\in[\nx]} \frac{1}{\sqrt{\Pr(a\vert x) \Pr^\rf(a\vert x)}}~ \sax\otimes (\sax^\rf)^\T,\\
& \sax\succeq 0, \qquad \forall a,x\\
& \tr(\sax E_{b\vert y}) = \Pr(a,b\vert x,y), \qquad \forall a,b,x,y,\\
& \Set{E_{b\vert y}}_{b}~~\ac{POVM},\qquad \forall y.
\end{aligned}
\end{equation}
Applying the relaxation described in the main text, i.e., the mapping in~\cref{Eq_Def_Gamma}, we obtain an \ac{SDP} of the form~\eqref{Eq_Fopt_final}:
\begin{equation}
\begin{aligned}
\text{given}\quad & \Pr(a,b\vert x,y),\Pr^\rf(a\vert x),\sax^\rf\\
\min_{\mathbf{u}} \quad & \Gamma(Y)_{\openone}\\
\text{subject to}\quad & \Gamma(Y)\otimes\openone\succeq \frac{1}{\nx} \sum_{a\in [\na],x\in [\nx]} \frac{1}{\sqrt{\Pr(a\vert x) \Pr^\rf(a\vert x)}}~\Gamma(\sax)\otimes (\sax^\rf)^\T,\\
&\Gamma(\sax)\succeq 0, \qquad \forall a,x,\\
&\Gamma(\sax)_{E_{b\vert y}} = \Pr(a,b\vert x,y), \qquad \forall a,b,x,y,
\end{aligned}
\end{equation}
where $\Gamma(\sax)=\sum_{i,j\in [m]}\ketbra{i}{j}\tr(S_j^\dag S_i \sax)$, with $\Set{S_i}_{i\in [m]}$ including the set of Bob's \ac{POVM} effects $\Set{E_{b\vert y}}_{b,y}$ and their products, were previously termed assemblage moment matrices in Ref.~\cite{Chen2021robustselftestingof}.  
In analogy to $\Gamma(Y)_{\openone}$ denoting the element of $\Gamma(Y)$ matrix corresponding to the moment $\tr(Y)$, by $\Gamma(\sax)_{E_{b\vert y}}$ we denote the entries of $\Gamma(\sax)$ corresponding to the moment $\tr(\sax E_{b\vert y})$.
The variables of optimization $\mathbf{u}$ are all entries of $\Gamma(\sax)$ and $\Gamma(Y)$ not fixed by $\Pr(a,b\vert x,y)$.

\subsection*{Example of CHSH scenario}
To plot~\cref{Fig_min_F_steering_CHSH}, we do not assume that the observed data is the conditional distribution $\Pr(a,b\vert x,y)$, but only the value of the \ac{CHSH} inequality $I_{\rm CHSH}$.
However, we do assume that $\Pr(a\vert x)=\frac{1}{2}$.
One can also obtain different bounds for other values of $\Pr(a\vert x)$.

As claimed in the text, the obtained lower bound on the fidelity is tight, i.e., there are quantum strategies that can reproduce the points on the line in \cref{Fig_min_F_steering_CHSH}.
First, observe that the point $(I_{\rm CHSH}, \Fid) = (2\sqrt{2}, 1)$ in~\cref{Fig_min_F_steering_CHSH} can be realized by the following assemblage $\sax$ of Alice and the observables of Bob $E_{0\vert y}-E_{1\vert y}$ of Bob, to which we refer as strategy $S_1$:
\begin{equation}
\Set{\sax}_{a,x\in\Set{0,1}} = \Set*{
\frac{\ketbra{0}{0}}{2},
\frac{\ketbra{1}{1}}{2},
\frac{\ketbra{+}{+}}{2},
\frac{\ketbra{-}{-}}{2}},\quad
\Set{E_{0\vert y}-E_{1\vert y}}_{y\in\Set{0,1}} = \Set*{
\frac{Z+X}{\sqrt{2}}, \frac{Z-X}{\sqrt{2}}}.
\end{equation}
Second, we observe that the point $(I_{\rm CHSH}, \Fid) = (2, 0.75)$ can be realized by the following strategy $S_2$:
\begin{equation}
\Set{\sax}_{a,x\in\Set{0,1}} = \Set*{
\frac{\ketbra{0}{0}}{2},
\frac{\ketbra{1}{1}}{2},
\frac{\openone}{4},
\frac{\openone}{4}},\quad
\{E_{b\vert y}\}_{b,y\in\Set{0,1}} = \Set*{
\ketbra{0}{0}, \ketbra{1}{1}, \ketbra{0}{0}, \ketbra{1}{1}}.
\end{equation}
With a simple \ac{SDP}, one can verify that for the strategy above, no higher value of the fidelity can be obtained.
Finally, with the convexity argument, all the points lying on the line between the above two points can be realized by convex mixtures of the strategies $S_1$ and $S_2$.

\section{Details of self-testing of constellation of states in the prepare-and-measure scenario}
\label{SecApp_SDP_PM}
We use the following definition of the averaged fidelity between two sets of qubit states, $\{\rx\}_{\vx\in [\nx]}$ and $\{\rx^\rf\}_{\vx\in [\nx]}$ (see e.g., \cite{SLChen2024,Tavakoli18b,SHWei2019,Miklin2021universalscheme}), up to equivalence transformation $\Lambda$:
\begin{equation}
\Fid = \frac{1}{2^n} \sum_{\vx\in[\nx]} \tr\left[\Lambda(\rx)\rx^\rf\right],
\end{equation}
where we assume that $\rx^\rf$ are pure for all $\vx\in[\nx]$. 
Compared with the general form of fidelity between general resources in \cref{Eq_Def_Fid}, we obtain the form of the set $I$, coefficients $c_i$, and states $\rho_i$ in~\cref{Eq_subst_2}.
With this, the optimization problem for finding the optimal fidelity~\eqref{Eq_Fopt_dual} is of the following form
\begin{equation}
\begin{aligned}
\Fid_\opt = \min_{\rx,Y,E_{b\vert y}} \quad & \tr(Y)\\
\text{subject to}\quad & Y\otimes\openone\succeq \frac{1}{2^n} \sum_{\vx\in[\nx]} \rx\otimes (\rx^\rf)^\T,\\
& \rx\succeq 0, \qquad \forall \vx\\
& \tr(\rx E_{b\vert y}) = \Pr(b\vert x,y), \qquad \forall b,\vx,y,\\
& \Set{E_{b\vert y}}_{b}~~\ac{POVM},\qquad \forall y.
\end{aligned}
\label{EqApp_SDP_PM}
\end{equation}
Applying the proposed \ac{SDP} relaxation approach, we obtain the optimization problem in the form of~\cref{Eq_Fopt_final}:
\begin{equation}
\begin{aligned}
\text{given}\quad & \Pr(b\vert \vx,y),\rx^\rf\\
\min_{\mathbf{u}} \quad & \Gamma(Y)_{\openone}\\
\text{subject to}\quad & \Gamma(Y)\otimes\openone\succeq \frac{1}{2^n} \sum_{\vx\in[\nx]} \Gamma(\rx)\otimes (\rx^\rf)^\T\\
&\Gamma(\rx)\succeq 0,\qquad\forall \vx,\\
&\Gamma(\rx)_{E_{b\vert y}} = \Pr(b\vert x,y), \qquad \forall b,\vx,y,
\end{aligned}
\end{equation}
where $\Gamma(\rx)=\sum_{i,j\in [m]}\ketbra{i}{j}\tr(S_j^\dag S_i \rx)$, with $\{S_i\}_{i\in[m]}$ containing \acp{POVM} effects of Bob $\Set{E_{b\vert y}}_{b,y}$ and their products, were previously termed the instrument moment matrices in Ref.~\cite{SLChen2024}.
The variables of optimization $\mathbf{u}$ include all entries of in $\Gamma(\rx)$ and $\Gamma(Y)$ not fixed by $\Pr(b\vert \vx,y)$. 
One important feature of self-testing in the prepare-and-measure scenario is that the dimension of $\HH$ (or equivalently, the communication between the parties) needs to be bounded. 
Such a condition can be incorporated as an additional constraint into the above \ac{SDP}~\cite{Navascues15prl,Navascues15,SLChen2024}:
\begin{equation}
\Gamma(\rx)\in\mathcal{G}_d,\qquad \forall \vx, 
\end{equation}
where $\mathcal{G}_d$ is the set of instrument moment matrices generated by $d$-dimensional systems~\cite{SLChen2024}. 
\subsection*{Example of $2\to 1$ RAC}
The observed data $\Pr(b\vert \vx,y)$ in \cref{EqApp_SDP_PM} can be substituted with a value of some figure of merit, e.g., of a $n\to 1$ \ac{RAC}, which is then the average guessing probability $\mathcal{P}_{n\to 1}=\frac{1}{n2^n}\sum_{\vx\in \Set{0,1}^n,y\in [n]} \Pr(b\vert \vx,y)\delta_{y,x_y}$, where $x_y$ is the $y$-th element of $\vx$.
In \cref{Fig_min_F_QRAC2to1} we show the results for $n=2$, which coincide with the previously reported values in Ref.~\cite{Tavakoli18b}, which, in the case of shared randomness, are tight. 

\section{Details of self-testing of entangled states in the steering scenario}
\label{SecApp_SDP_ST_1SDI}
As mentioned in the main text, for self-testing of entangled states $\rab\in \DM(\HA\otimes\HB)$ in the steering scenario, we have an additional restriction that the equivalence map $\Lambda$ only acts on Alice's subsystem, i.e., is local~\cite{Supic2020selftestingof}.
For this reason, we give a more detailed derivation here, although the general structure of the dual optimization problem and the \ac{SDP} relaxation still have the form as in \cref{Eq_Fopt_dual} and \cref{Eq_Fopt_final}.

We start by writing the formula for the fidelity of the physical and pure target entangled states up to equivalence \ac{CPTP} map $\Lambda:\LL(\HH_\A)\to\LL(\HH_{\A'})$,
\begin{equation}
    \Fid = \tr\left[\rho^\rf_{\A\B}(\Lambda\otimes\id)(\rab)\right] = \tr(\rho_{\A\B}^\T\otimes\rho_{\A'\B'}^\rf~\Omega_{\A\A'}\otimes{\kbpsi}_{\B\B'}) = \tr\left(\rab\otimes (\rho_{\A'\B'}^\rf)^\T~\Omega^\T_{\A\A'}\otimes{\kbpsi}_{\B\B'}\right),
\end{equation}
where, as before $\Omega = (\id\otimes\Lambda)(\kbpsi)$, and, with a slight abuse of notation, we denote by $\ket{\psi^+} = \sum_{i}\ket{i}\otimes\ket{i}$ the unnormalized maximally entangled state in both $\HA\otimes\HH_{\A'}$ and $\HB\otimes\HH_{\B'}$.

Let us fix $\rho^\ast_{\A\B}$ to be a state that is compatible with the observed data $\Set{\sax}_{a\in[\na],x\in[\nx]}$.
The optimization problem with respect to $\Omega$ is then the following
\begin{equation}
\begin{aligned}
\text{given} \quad &\rho^\ast_{\A\B}, \rho^\rf_{\A'\B'}\\
\max\quad & \tr\left(\rho^\ast_{\A\B}\otimes(\rho_{\A'\B'}^\rf)^\T~\Omega_{\A\A'}\otimes{\kbpsi}_{\B\B'}\right)\\
\text{subject to}\quad &\Omega\succeq 0,\quad \tr_{\A'}(\Omega) = \openone,
\end{aligned}
\label{EqApp_Max_1SDI}
\end{equation}
where, again, we changed the optimization variable from $\Omega$ to $\Omega^\T$.
Comparing \cref{EqApp_Max_1SDI} with \cref{EqApp_max_over_Omega}, we conclude that the dual \ac{SDP} to \eqref{EqApp_Max_1SDI} is
\begin{equation}
\begin{aligned}
    \min_{Y}\quad & \tr(Y)\\
    \text{subject to}\quad & Y\otimes \openone \succeq \tr_{\B\B'}\left(\rho^\ast_{\A\B}\otimes(\rho_{\A'\B'}^\rf)^\T~\openone_{\A\A'}\otimes{\kbpsi}_{\B\B'}\right).
\end{aligned}
\end{equation}
Returning the minimization over $\rab$, and adding the constraint of compatibility with the observed data, we obtain the following optimization problem,
\begin{equation}
\begin{aligned}
    \text{given}\quad & \sax, \rho^\rf_{\A'\B'}\\
    \min_{Y,\rab,E_{a\vert x}}\quad & \tr(Y)\\
    \text{subject to}\quad & Y\otimes \openone \succeq \tr_{\B\B'}\left(\rab\otimes(\rho_{\A'\B'}^\rf)^\T~\openone_{\A\A'}\otimes{\kbpsi}_{\B\B'}\right),\\
    & \rab\succeq 0,\\
    & \tr_\A(E_{a\vert x}\otimes \openone\rab) = \sax, \qquad \forall a,x,\\
    & \Set{E_{a\vert x}}_a~~\ac{POVM}, \qquad \forall x.
\end{aligned}
\end{equation}
Now, we apply the \ac{SDP} relaxation map~\eqref{Eq_Def_Gamma}.
Importantly, we apply it only to the subsystem $\A$, which results in the following \ac{SDP},
\begin{equation}
\begin{aligned}
    \text{given}\quad & \sax, \rho^\rf_{\A'\B'}\\
    \min_{\bm{u}}\quad & \Gamma(Y)_{\openone}\\
    \text{subject to}\quad & \Gamma(Y)\otimes \openone \succeq \tr_{\B\B'}\left((\Gamma\otimes\id)(\rab)\otimes(\rho_{\A'\B'}^\rf)^\T~\openone_{\A\A'}\otimes{\kbpsi}_{\B\B'}\right),\\
    & (\Gamma\otimes\id)(\rab)\succeq 0,\\
    & (\Gamma\otimes\id)(\rab)_{E_{a\vert x}} = \sax, \qquad \forall a,x,\\
\label{EqApp_SDP_SDI_ST_assemblage}
\end{aligned}
\end{equation}
where $(\Gamma\otimes\id)(\rab) = \sum_{i,j\in [m]}\ketbra{i}{j}\otimes\tr_{\A}\left(S_j^\dagger S_i\otimes\openone\rab\right)$, and $\Gamma(Y) = \sum_{i,j\in [m]}\ketbra{i}{j}\tr\left(S_j^\dagger S_iY\right)$, for the set of operators $\Set{S_i}_{i\in[m]}$, which include the effects of Alice's \acp{POVM} $\Set{E_{a\vert x}}_{a,x}$ and their products (see also Ref.~\cite{Pusey13}). 


\subsection*{Example of the simplest steering scenario}
As an example, we again consider a scenario with binary measurement settings and outcomes on Alice's side.
To plot the robustness bounds, we once again substitute the observed assemblage $\Set{\sax}_{a\in[\na],x\in[\nx]}$ in~\cref{EqApp_SDP_SDI_ST_assemblage} with an observed violation of the steering inequality $I_{\rm steering} = \tr\left[Z(\sigma_{0|0}-\sigma_{1|0}) + X(\sigma_{0|1}-\sigma_{1|1})\right]$.
We show our results in~\cref{Fig_minF_1SDI} and compare them with those of Ref.~\cite{Supic2016}.

\twocolumngrid
\bibliography{bib_self_testing}

@inproceedings{Lofberg2004,
address = {Taipei, Taiwan},
author = {L{\"{o}}fberg, J.},
booktitle = {In Proceedings of the CACSD Conference},
title = {{YALMIP}: A Toolbox for Modeling and Optimization in {MATLAB}},
doi = {10.1109/CACSD.2004.1393890},
year = {2004}}

@article{SDPT3,
author = {Toh, K. C. and Todd, M. J. and Tütüncü, R. H.},
title = {{SDPT3} — A {MATLAB} software package for semidefinite programming, Version 1.3},
journal = {Optimization Methods and Software},
volume = {11},
number = {1-4},
pages = {545--581},
year = {1999},
publisher = {Taylor \& Francis},
doi = {10.1080/10556789908805762},
}

@book{Skrzypczyk2023SDP,
author = {Skrzypczyk, Paul and Cavalcanti, Daniel},
title = {Semidefinite Programming in Quantum Information Science},
publisher = {IOP Publishing},
year = {2023},
series = {2053-2563},
isbn = {978-0-7503-3343-6},
url = {https://dx.doi.org/10.1088/978-0-7503-3343-6},
doi = {10.1088/978-0-7503-3343-6}
}

@inproceedings{Ambainis1999,
author = {Ambainis, Andris and Nayak, Ashwin and Ta-Shma, Ammon and Vazirani, Umesh},
title = {Dense quantum coding and a lower bound for 1-way quantum automata},
year = {1999},
isbn = {1581130678},
publisher = {Association for Computing Machinery},
address = {New York, NY, USA},
url = {https://doi.org/10.1145/301250.301347},
doi = {10.1145/301250.301347},
booktitle = {Proceedings of the Thirty-First Annual ACM Symposium on Theory of Computing},
pages = {376–383},
numpages = {8},
location = {Atlanta, Georgia, USA},
series = {STOC '99}
}

@article{Tavakoli2023Review,
  title = {Semidefinite programming relaxations for quantum correlations},
  author = {Tavakoli, Armin and Pozas-Kerstjens, Alejandro and Brown, Peter and Ara\'ujo, Mateus},
  journal = {Rev. Mod. Phys.},
  volume = {96},
  issue = {4},
  pages = {045006},
  numpages = {68},
  year = {2024},
  month = {Dec},
  publisher = {American Physical Society},
  doi = {10.1103/RevModPhys.96.045006},
  url = {https://link.aps.org/doi/10.1103/RevModPhys.96.045006}
}

@article{Supic2016,
  title = {Self-testing through {EPR}-steering},
  volume = {18},
  ISSN = {1367-2630},
  url = {http://dx.doi.org/10.1088/1367-2630/18/7/075006},
  DOI = {10.1088/1367-2630/18/7/075006},
  number = {7},
  journal = {New Journal of Physics},
  publisher = {IOP Publishing},
  author = {{\v{S}}upi{\'{c}},  Ivan and Hoban,  Matty J},
  year = {2016},
  month = July,
  pages = {075006}
}

@article{Tavakoli2020self,
  title = {Self-testing nonprojective quantum measurements in prepare-and-measure experiments},
  volume = {6},
  ISSN = {2375-2548},
  url = {http://dx.doi.org/10.1126/sciadv.aaw6664},
  DOI = {10.1126/sciadv.aaw6664},
  number = {16},
  journal = {Science Advances},
  publisher = {American Association for the Advancement of Science (AAAS)},
  author = {Tavakoli,  Armin and Smania,  Massimiliano and Vértesi,  Tamás and Brunner,  Nicolas and Bourennane,  Mohamed},
  year = {2020},
  month = apr 
}

@article{Kliesch2021theory,
  title = {Theory of Quantum System Certification},
  author = {Kliesch, Martin and Roth, Ingo},
  journal = {PRX Quantum},
  volume = {2},
  issue = {1},
  pages = {010201},
  numpages = {53},
  year = {2021},
  month = {Jan},
  publisher = {American Physical Society},
  doi = {10.1103/PRXQuantum.2.010201},
  url = {https://link.aps.org/doi/10.1103/PRXQuantum.2.010201}
}

@article{Miklin2020,
  title = {Semi-device-independent self-testing of unsharp measurements},
  author = {Miklin, Nikolai and Borka\l{}a, Jakub J. and Paw\l{}owski, Marcin},
  journal = {Phys. Rev. Res.},
  volume = {2},
  issue = {3},
  pages = {033014},
  numpages = {15},
  year = {2020},
  month = {Jul},
  publisher = {American Physical Society},
  doi = {10.1103/PhysRevResearch.2.033014},
  url = {https://link.aps.org/doi/10.1103/PhysRevResearch.2.033014}
}

@article{Nadlinger2022,
  title = {Experimental quantum key distribution certified by {B}ell's theorem},
  volume = {607},
  ISSN = {1476-4687},
  url = {http://dx.doi.org/10.1038/s41586-022-04941-5},
  DOI = {10.1038/s41586-022-04941-5},
  number = {7920},
  journal = {Nature},
  publisher = {Springer Science and Business Media LLC},
  author = {Nadlinger,  D. P. and Drmota,  P. and Nichol,  B. C. and Araneda,  G. and Main,  D. and Srinivas,  R. and Lucas,  D. M. and Ballance,  C. J. and Ivanov,  K. and Tan,  E. Y.-Z. and Sekatski,  P. and Urbanke,  R. L. and Renner,  R. and Sangouard,  N. and Bancal,  J.-D.},
  year = {2022},
  month = jul,
  pages = {682–686}
}

@article{Eisert2020,
  title = {Quantum certification and benchmarking},
  volume = {2},
  ISSN = {2522-5820},
  url = {http://dx.doi.org/10.1038/s42254-020-0186-4},
  DOI = {10.1038/s42254-020-0186-4},
  number = {7},
  journal = {Nature Reviews Physics},
  publisher = {Springer Science and Business Media LLC},
  author = {Eisert,  Jens and Hangleiter,  Dominik and Walk,  Nathan and Roth,  Ingo and Markham,  Damian and Parekh,  Rhea and Chabaud,  Ulysse and Kashefi,  Elham},
  year = {2020},
  month = jun,
  pages = {382–390}
}

@article{SLChen2024,
  title = {Semi-Device-Independently Characterizing Quantum Temporal Correlations},
  author = {Chen, Shin-Liang and Eisert, Jens},
  journal = {Phys. Rev. Lett.},
  volume = {132},
  issue = {22},
  pages = {220201},
  numpages = {7},
  year = {2024},
  month = {May},
  publisher = {American Physical Society},
  doi = {10.1103/PhysRevLett.132.220201},
  url = {https://link.aps.org/doi/10.1103/PhysRevLett.132.220201}
}

@article{Mohan2019,
doi = {10.1088/1367-2630/ab3773},
url = {https://doi.org/10.1088/1367-2630/ab3773},
year = {2019},
month = {aug},
publisher = {IOP Publishing},
volume = {21},
number = {8},
pages = {083034},
author = {Mohan, Karthik and Tavakoli, Armin and Brunner, Nicolas},
title = {Sequential random access codes and self-testing of quantum measurement instruments},
journal = {New Journal of Physics},
}

@article{Anwer2020,
  title = {Experimental Characterization of Unsharp Qubit Observables and Sequential Measurement Incompatibility via Quantum Random Access Codes},
  author = {Anwer, Hammad and Muhammad, Sadiq and Cherifi, Walid and Miklin, Nikolai and Tavakoli, Armin and Bourennane, Mohamed},
  journal = {Phys. Rev. Lett.},
  volume = {125},
  issue = {8},
  pages = {080403},
  numpages = {7},
  year = {2020},
  month = {Aug},
  publisher = {American Physical Society},
  doi = {10.1103/PhysRevLett.125.080403},
  url = {https://link.aps.org/doi/10.1103/PhysRevLett.125.080403}
}

@article{Kaniewski2017,
  title = {Self-testing of binary observables based on commutation},
  author = {Kaniewski, J{\k{e}}drzej},
  journal = {Phys. Rev. A},
  volume = {95},
  issue = {6},
  pages = {062323},
  numpages = {10},
  year = {2017},
  month = {Jun},
  publisher = {American Physical Society},
  doi = {10.1103/PhysRevA.95.062323},
  url = {https://link.aps.org/doi/10.1103/PhysRevA.95.062323}
}

@article{Pawowski2011,
  title = {Semi-device-independent security of one-way quantum key distribution},
  author = {Paw\l{}owski, Marcin and Brunner, Nicolas},
  journal = {Phys. Rev. A},
  volume = {84},
  issue = {1},
  pages = {010302},
  numpages = {4},
  year = {2011},
  month = {Jul},
  publisher = {American Physical Society},
  doi = {10.1103/PhysRevA.84.010302},
  url = {https://link.aps.org/doi/10.1103/PhysRevA.84.010302}
}

@article{Farkas19,
  title = {Self-testing mutually unbiased bases in the prepare-and-measure scenario},
  author = {Farkas, M\'at\'e and Kaniewski, J{\k{e}}drzej},
  journal = {Phys. Rev. A},
  volume = {99},
  issue = {3},
  pages = {032316},
  numpages = {11},
  year = {2019},
  month = {Mar},
  publisher = {American Physical Society},
  doi = {10.1103/PhysRevA.99.032316},
  url = {https://link.aps.org/doi/10.1103/PhysRevA.99.032316}
}

@article{Navascues2023,
  title = {Self-Testing in Prepare-and-Measure Scenarios and a Robust Version of {W}igner's Theorem},
  author = {Navascu\'es, Miguel and P\'al, K\'aroly F. and V\'ertesi, Tam\'as and Ara\'ujo, Mateus},
  journal = {Phys. Rev. Lett.},
  volume = {131},
  issue = {25},
  pages = {250802},
  numpages = {6},
  year = {2023},
  month = {Dec},
  publisher = {American Physical Society},
  doi = {10.1103/PhysRevLett.131.250802},
  url = {https://link.aps.org/doi/10.1103/PhysRevLett.131.250802}
}

@article{Supic2020selftestingof,
  doi = {10.22331/q-2020-09-30-337},
  url = {https://doi.org/10.22331/q-2020-09-30-337},
  title = {Self-testing of quantum systems: a review},
  author = {{\v{S}}upi{\'{c}}, Ivan and Bowles, Joseph},
  journal = {{Quantum}},
  issn = {2521-327X},
  publisher = {{Verein zur F{\"{o}}rderung des Open Access Publizierens in den Quantenwissenschaften}},
  volume = {4},
  pages = {337},
  month = sep,
  year = {2020}
}

@article{SHWei2019,
	doi = {10.1088/1674-1056/28/7/070304},
	url = {https://doi.org/10.1088/1674-1056/28/7/070304},
	year = 2019,
	month = {jul},
	publisher = {{IOP} Publishing},
	volume = {28},
	number = {7},
	pages = {070304},
	author = {Shi-Hui Wei and Fen-Zhuo Guo and Xin-Hui Li and Qiao-Yan Wen},
	title = {Robustness self-testing of states and measurements in the prepare-and-measure scenario with $3\rightarrow 1$ random access code},
	journal = {Chinese Phys. B}
}

@article{Miklin2021universalscheme,
  doi = {10.22331/q-2021-04-06-424},
  url = {https://doi.org/10.22331/q-2021-04-06-424},
  title = {A universal scheme for robust self-testing in the prepare-and-measure scenario},
  author = {Miklin, Nikolai and Oszmaniec, Micha{\l{}}},
  journal = {{Quantum}},
  issn = {2521-327X},
  publisher = {{Verein zur F{\"{o}}rderung des Open Access Publizierens in den Quantenwissenschaften}},
  volume = {5},
  pages = {424},
  month = apr,
  year = {2021}
}

@article{Tavakoli18b,
  title = {Self-testing quantum states and measurements in the prepare-and-measure scenario},
  author = {Tavakoli, Armin and Kaniewski, J{\k{e}}drzej and V\'ertesi, Tam\'as and Rosset, Denis and Brunner, Nicolas},
  journal = {Phys. Rev. A},
  volume = {98},
  issue = {6},
  pages = {062307},
  numpages = {13},
  year = {2018},
  month = {Dec},
  publisher = {American Physical Society},
  doi = {10.1103/PhysRevA.98.062307},
  url = {https://link.aps.org/doi/10.1103/PhysRevA.98.062307}
}

@article{CMBC2021,
  title = {Device-independent quantification of measurement incompatibility},
  author = {Chen, Shin-Liang and Miklin, Nikolai and Budroni, Costantino and Chen, Yueh-Nan},
  journal = {Phys. Rev. Research},
  volume = {3},
  issue = {2},
  pages = {023143},
  numpages = {15},
  year = {2021},
  month = {May},
  publisher = {American Physical Society},
  doi = {10.1103/PhysRevResearch.3.023143},
  url = {https://link.aps.org/doi/10.1103/PhysRevResearch.3.023143}
}

@article{Uola2020Steering,
  title = {Quantum steering},
  author = {Uola, Roope and Costa, Ana C. S. and Nguyen, H. Chau and G\"uhne, Otfried},
  journal = {Rev. Mod. Phys.},
  volume = {92},
  issue = {1},
  pages = {015001},
  numpages = {40},
  year = {2020},
  month = {Mar},
  publisher = {American Physical Society},
  doi = {10.1103/RevModPhys.92.015001},
  url = {https://link.aps.org/doi/10.1103/RevModPhys.92.015001}
}

@article{Tavakoli2020Semi,
  title = {Semi-Device-Independent Certification of Independent Quantum State and Measurement Devices},
  author = {Tavakoli, Armin},
  journal = {Phys. Rev. Lett.},
  volume = {125},
  issue = {15},
  pages = {150503},
  numpages = {6},
  year = {2020},
  month = {Oct},
  publisher = {American Physical Society},
  doi = {10.1103/PhysRevLett.125.150503},
  url = {https://link.aps.org/doi/10.1103/PhysRevLett.125.150503}
}

@article{Chen2021robustselftestingof,
  doi = {10.22331/q-2021-09-28-552},
  url = {https://doi.org/10.22331/q-2021-09-28-552},
  title = {Robust self-testing of steerable quantum assemblages and its applications on device-independent quantum certification},
  author = {Chen, Shin-Liang and Ku, Huan-Yu and Zhou, Wenbin and Tura, Jordi and Chen, Yueh-Nan},
  journal = {{Quantum}},
  issn = {2521-327X},
  publisher = {{Verein zur F{\"{o}}rderung des Open Access Publizierens in den Quantenwissenschaften}},
  volume = {5},
  pages = {552},
  month = sep,
  year = {2021}
}

@article{Noller2025classical,
  doi = {10.22331/q-2025-08-08-1825},
  url = {https://doi.org/10.22331/q-2025-08-08-1825},
  title = {Classical certification of quantum gates under the dimension assumption},
  author = {N{\"{o}}ller, Jan and Miklin, Nikolai and Kliesch, Martin and Gachechiladze, Mariami},
  journal = {{Quantum}},
  issn = {2521-327X},
  publisher = {{Verein zur F{\"{o}}rderung des Open Access Publizierens in den Quantenwissenschaften}},
  volume = {9},
  pages = {1825},
  month = aug,
  year = {2025}
}

@book{wigner1931gruppentheorie,
  title = {Gruppentheorie und ihre Anwendung auf die Quantenmechanik der Atomspektren},
  ISBN = {9783663025559},
  url = {http://dx.doi.org/10.1007/978-3-663-02555-9},
  DOI = {10.1007/978-3-663-02555-9},
  publisher = {Vieweg+Teubner Verlag},
  author = {Wigner,  Eugen},
  year = {1931}
}

@article{Noller2025sound,
      title={Sound certification of memory-bounded quantum computers}, 
      author={Jan Nöller and Nikolai Miklin and Martin Kliesch and Mariami Gachechiladze},
      year={2025},
      eprint={2411.04215},
      archivePrefix={arXiv},
      primaryClass={quant-ph},
      url={https://arxiv.org/abs/2411.04215}, 
}

@article{Gallego10,
  title = {Device-Independent Tests of Classical and Quantum Dimensions},
  author = {Gallego, Rodrigo and Brunner, Nicolas and Hadley, Christopher and Ac\'{\i}n, Antonio},
  journal = {Phys. Rev. Lett.},
  volume = {105},
  issue = {23},
  pages = {230501},
  numpages = {4},
  year = {2010},
  month = {Nov},
  publisher = {American Physical Society},
  doi = {10.1103/PhysRevLett.105.230501},
  url = {https://link.aps.org/doi/10.1103/PhysRevLett.105.230501}
}

@article{Pironio10b,
  doi = {10.1137/090760155},
  url = {https://doi.org/10.1137/090760155},
  year = {2010},
  month = jan,
  publisher = {Society for Industrial {\&} Applied Mathematics ({SIAM})},
  volume = {20},
  number = {5},
  pages = {2157--2180},
  author = {S. Pironio and M. Navascu{\'{e}}s and A. Ac{\'{\i}}n},
  title = {Convergent Relaxations of Polynomial Optimization Problems with Noncommuting Variables},
  journal = {{SIAM} Journal on Optimization}
}

@article{Liang19,
  doi = {10.1088/1361-6633/ab1ca4},
  url = {https://doi.org/10.1088/1361-6633/ab1ca4},
  year = {2019},
  month = jun,
  publisher = {{IOP} Publishing},
  volume = {82},
  number = {7},
  pages = {076001},
  author = {Yeong-Cherng Liang and Yu-Hao Yeh and Paulo E M F Mendon{\c{c}}a and Run Yan Teh and Margaret D Reid and Peter D Drummond},
  title = {Quantum fidelity measures for mixed states},
  journal = {Rep. Prof. Phys.}
}

@article{CBLC18,
  title = {Exploring the framework of assemblage moment matrices and its applications in device-independent characterizations},
  author = {Chen, Shin-Liang and Budroni, Costantino and Liang, Yeong-Cherng and Chen, Yueh-Nan},
  journal = {Phys. Rev. A},
  volume = {98},
  issue = {4},
  pages = {042127},
  numpages = {15},
  year = {2018},
  month = {Oct},
  publisher = {American Physical Society},
  doi = {10.1103/PhysRevA.98.042127},
  url = {https://link.aps.org/doi/10.1103/PhysRevA.98.042127}
}

@article{CBLC16,
  title = {Natural Framework for Device-Independent Quantification of Quantum Steerability, Measurement Incompatibility, and Self-Testing},
  author = {Chen, Shin-Liang and Budroni, Costantino and Liang, Yeong-Cherng and Chen, Yueh-Nan},
  journal = {Phys. Rev. Lett.},
  volume = {116},
  issue = {24},
  pages = {240401},
  numpages = {7},
  year = {2016},
  month = {Jun},
  publisher = {American Physical Society},
  doi = {10.1103/PhysRevLett.116.240401},
  url = {https://link.aps.org/doi/10.1103/PhysRevLett.116.240401}
}

@article{Kaniewski16,
  title = {Analytic and Nearly Optimal Self-Testing Bounds for the {C}lauser-{H}orne-{S}himony-{H}olt and {M}ermin Inequalities},
  author = {Kaniewski, J\k{e}drzej},
  journal = {Phys. Rev. Lett.},
  volume = {117},
  issue = {7},
  pages = {070402},
  numpages = {6},
  year = {2016},
  month = {Aug},
  publisher = {American Physical Society},
  doi = {10.1103/PhysRevLett.117.070402},
  url = {https://link.aps.org/doi/10.1103/PhysRevLett.117.070402}
}

@article{Tsirelson93,
journal = "Hadronic Journal Supplement",
title = "Some results and problems on quantum {B}ell-type inequalities",
volume = "8",
number = "",
pages = "329 - 345",
year = "1993",
url = "https://ma.huji.ac.il/~ohadfeld/Tsirelson/download/hadron.html",
author = "B. S. Tsirelson"
}

@article{Popescu92,
title = "Which states violate {B}ell's inequality maximally?",
journal = "Physics Letters A",
volume = "169",
number = "6",
pages = "411 - 414",
year = "1992",
issn = "0375-9601",
doi = "https://doi.org/10.1016/0375-9601(92)90819-8",
url = "http://www.sciencedirect.com/science/article/pii/0375960192908198",
author = "Sandu Popescu and Daniel Rohrlich"
}

@article{Summers87,
author = "Summers, Stephen J. and Werner, Reinhard",
fjournal = "Communications in Mathematical Physics",
journal = "Comm. Math. Phys.",
number = "2",
pages = "247--259",
publisher = "Springer",
title = "Maximal violation of {B}ell's inequalities is generic in quantum field theory",
url = "https://projecteuclid.org:443/euclid.cmp/1104159237",
volume = "110",
year = "1987"
}

@article{Navascues15prl,
	Author = {Navascu\'es, Miguel and V\'ertesi, Tam\'as},
	Doi = {10.1103/PhysRevLett.115.020501},
	Issue = {2},
	Journal = {Phys. Rev. Lett.},
	Month = {Jul},
	Numpages = {5},
	Pages = {020501},
	Publisher = {American Physical Society},
	Title = {Bounding the Set of Finite Dimensional Quantum Correlations},
	Url = {https://link.aps.org/doi/10.1103/PhysRevLett.115.020501},
	Volume = {115},
	Year = {2015}}

@article{Pusey13,
	Author = {Pusey, Matthew F.},
	Doi = {10.1103/PhysRevA.88.032313},
	Issue = {3},
	Journal = {Phys. Rev. A},
	Month = {Sep},
	Numpages = {5},
	Pages = {032313},
	Publisher = {American Physical Society},
	Title = {Negativity and steering: A stronger {P}eres conjecture},
	Url = {https://link.aps.org/doi/10.1103/PhysRevA.88.032313},
	Volume = {88},
	Year = {2013}}

@book{BoydBook,
	Author = {Stephen Boyd AND Lieven Vandenberghe},
	Edition = {1},
	Publisher = {Cambridge University Press, Cambridge},
	Title = {Convex optimization},
    doi = {10.1017/CBO9780511804441},
	Year = {2004}}

@article{Choi75,
	Author = {Choi, Man-Duen},
	Da = {1975/06/01},
	Doi = {http://dx.doi.org/10.1016/0024-3795(75)90075-0},
	Isbn = {0024-3795},
	Journal = {Linear Algebra Appl.},
	Number = {3},
	Pages = {285--290},
	Title = {Completely positive linear maps on complex matrices},
	Ty = {JOUR},
	Url = {http://www.sciencedirect.com/science/article/pii/0024379575900750},
	Volume = {10},
	Year = {1975}}

@article{Jamiokowski74,
	Author = {Jamio{\l}kowski, A.},
	Date = {1974/6//},
	Doi = {http://dx.doi.org/10.1016/0034-4877(74)90044-5},
	Isbn = {0034-4877},
	Journal = {Rep. Math. Phys.},
	Month = {6},
	Number = {3},
	Pages = {415--424},
	Title = {An effective method of investigation of positive maps on the set of positive definite operators},
	Ty = {JOUR},
	Url = {http://www.sciencedirect.com/science/article/pii/0034487774900445},
	Volume = {5},
	Year = {1974}}

@article{NPA,
	Author = {Navascu\'es, Miguel and Pironio, Stefano and Ac\'{\i}n, Antonio},
	Doi = {10.1103/PhysRevLett.98.010401},
	Issue = {1},
	Journal = {Phys. Rev. Lett.},
	Month = {Jan},
	Numpages = {4},
	Pages = {010401},
	Publisher = {American Physical Society},
	Title = {Bounding the Set of Quantum Correlations},
	Url = {http://link.aps.org/doi/10.1103/PhysRevLett.98.010401},
	Volume = {98},
	Year = {2007}}

@article{Kogias15,
	Author = {Kogias, Ioannis and Skrzypczyk, Paul and Cavalcanti, Daniel and Ac\'{\i}n, Antonio and Adesso, Gerardo},
	Doi = {10.1103/PhysRevLett.115.210401},
	Issue = {21},
	Journal = {Phys. Rev. Lett.},
	Month = {Nov},
	Numpages = {7},
	Pages = {210401},
	Publisher = {American Physical Society},
	Title = {Hierarchy of Steering Criteria Based on Moments for All Bipartite Quantum Systems},
	Url = {http://link.aps.org/doi/10.1103/PhysRevLett.115.210401},
	Volume = {115},
	Year = {2015}}

@article{Wolf09,
	Author = {Wolf, Michael M. and Perez-Garcia, David and Fernandez, Carlos},
	Doi = {10.1103/PhysRevLett.103.230402},
	Issue = {23},
	Journal = {Phys. Rev. Lett.},
	Month = {Dec},
	Numpages = {4},
	Pages = {230402},
	Publisher = {American Physical Society},
	Title = {Measurements Incompatible in Quantum Theory Cannot Be Measured Jointly in Any Other No-Signaling Theory},
	Url = {http://link.aps.org/doi/10.1103/PhysRevLett.103.230402},
	Volume = {103},
	Year = {2009}}

@article{Wiseman07,
	Author = {Wiseman, H. M. and Jones, S. J. and Doherty, A. C.},
	Doi = {10.1103/PhysRevLett.98.140402},
	Issue = {14},
	Journal = {Phys. Rev. Lett.},
	Month = {Apr},
	Numpages = {4},
	Pages = {140402},
	Publisher = {American Physical Society},
	Title = {Steering, Entanglement, Nonlocality, and the {E}instein-{P}odolsky-{R}osen Paradox},
	Url = {http://link.aps.org/doi/10.1103/PhysRevLett.98.140402},
	Volume = {98},
	Year = {2007}}

@article{Bell64,
  title = {On the {E}instein {P}odolsky {R}osen paradox},
  author = {Bell, J. S.},
  journal = {Physics Physique Fizika},
  volume = {1},
  issue = {3},
  pages = {195--200},
  numpages = {6},
  year = {1964},
  month = {Nov},
  publisher = {American Physical Society},
  doi = {10.1103/PhysicsPhysiqueFizika.1.195},
  url = {https://link.aps.org/doi/10.1103/PhysicsPhysiqueFizika.1.195}
}

@article{Acin07,
	Author = {Ac\'{\i}n, Antonio and Brunner, Nicolas and Gisin, Nicolas and Massar, Serge and Pironio, Stefano and Scarani, Valerio},
	Doi = {10.1103/PhysRevLett.98.230501},
	Issue = {23},
	Journal = {Phys. Rev. Lett.},
	Month = {Jun},
	Numpages = {4},
	Pages = {230501},
	Publisher = {American Physical Society},
	Title = {Device-Independent Security of Quantum Cryptography against Collective Attacks},
	Url = {http://link.aps.org/doi/10.1103/PhysRevLett.98.230501},
	Volume = {98},
	Year = {2007}}

@article{Mayers04,
	Acmid = {2011830},
	Author = {Mayers, Dominic and Yao, Andrew},
	Issn = {1533-7146},
	Issue_Date = {July 2004},
	Journal = {Quantum Info. Comput.},
	Month = jul,
	Number = {4},
	Numpages = {14},
	Pages = {273},
	Publisher = {Rinton Press, Incorporated},
	Title = {Self Testing Quantum Apparatus},
	Url = {http://dl.acm.org/citation.cfm?id=2011827.2011830},
	Volume = {4},
	Year = {2004}}

@article{McKague12_0,
	Author = {McKague, M. and Yang, T. H. and Scarani, V.},
	Journal = {Journal of Physics A: Mathematical and Theoretical},
	Number = {45},
	Pages = {455304},
	Title = {Robust self-testing of the singlet},
	Url = {http://stacks.iop.org/1751-8121/45/i=45/a=455304},
	Volume = {45},
	Year = {2012}}

@article{Yang14,
	Author = {Yang, Tzyh Haur and V\'ertesi, Tam\'as and Bancal, Jean-Daniel and Scarani, Valerio and Navascu\'es, Miguel},
	Doi = {10.1103/PhysRevLett.113.040401},
	Issue = {4},
	Journal = {Phys. Rev. Lett.},
	Month = {Jul},
	Numpages = {5},
	Pages = {040401},
	Publisher = {American Physical Society},
	Title = {Robust and Versatile Black-Box Certification of Quantum Devices},
	Url = {http://link.aps.org/doi/10.1103/PhysRevLett.113.040401},
	Volume = {113},
	Year = {2014}}

@article{Bancal15,
	Author = {Bancal, Jean-Daniel and Navascu\'es, Miguel and Scarani, Valerio and V\'ertesi, Tam\'as and Yang, Tzyh Haur},
	Doi = {10.1103/PhysRevA.91.022115},
	Issue = {2},
	Journal = {Phys. Rev. A},
	Month = {Feb},
	Numpages = {16},
	Pages = {022115},
	Publisher = {American Physical Society},
	Title = {Physical characterization of devices from nonlocal correlations},
	Url = {http://link.aps.org/doi/10.1103/PhysRevA.91.022115},
	Volume = {91},
	Year = {2015}}

@article{Pironio10,
  doi = {10.1038/nature09008},
  url = {https://doi.org/10.1038/nature09008},
  year = {2010},
  month = apr,
  publisher = {Springer Science and Business Media {LLC}},
  volume = {464},
  number = {7291},
  pages = {1021--1024},
  author = {S. Pironio and A. Ac{\'{\i}}n and S. Massar and A. Boyer de la Giroday and D. N. Matsukevich and P. Maunz and S. Olmschenk and D. Hayes and L. Luo and T. A. Manning and C. Monroe},
  title = {Random numbers certified by {B}ell's theorem},
  journal = {Nature}
}

@article{Navascues15,
	Author = {Navascu\'es, Miguel and Feix, Adrien and Ara\'ujo, Mateus and V\'ertesi, Tam\'as},
	Doi = {10.1103/PhysRevA.92.042117},
	Issue = {4},
	Journal = {Phys. Rev. A},
	Month = {Oct},
	Numpages = {15},
	Pages = {042117},
	Publisher = {American Physical Society},
	Title = {Characterizing finite-dimensional quantum behavior},
	Url = {http://link.aps.org/doi/10.1103/PhysRevA.92.042117},
	Volume = {92},
	Year = {2015}}

@article{Moroder13,
	Author = {Moroder, Tobias and Bancal, Jean-Daniel and Liang, Yeong-Cherng and Hofmann, Martin and G\"uhne, Otfried},
	Doi = {10.1103/PhysRevLett.111.030501},
	Issue = {3},
	Journal = {Phys. Rev. Lett.},
	Month = {Jul},
	Numpages = {5},
	Pages = {030501},
	Publisher = {American Physical Society},
	Title = {Device-Independent Entanglement Quantification and Related Applications},
	Url = {http://link.aps.org/doi/10.1103/PhysRevLett.111.030501},
	Volume = {111},
	Year = {2013}}

@article{Clauser69,
	Author = {Clauser, John F. and Horne, Michael A. and Shimony, Abner and Holt, Richard A.},
	Doi = {10.1103/PhysRevLett.23.880},
	Issue = {15},
	Journal = {Phys. Rev. Lett.},
	Month = {Oct},
	Numpages = {0},
	Pages = {880--884},
	Publisher = {American Physical Society},
	Title = {Proposed Experiment to Test Local Hidden-Variable Theories},
	Url = {http://link.aps.org/doi/10.1103/PhysRevLett.23.880},
	Volume = {23},
	Year = {1969}}

@article{NPA2008,
	Author = {Miguel Navascu{\'e}s and Stefano Pironio and Antonio Ac{\'\i}n},
	Journal = {New J. Phys.},
	Number = {7},
	Pages = {073013},
	Title = {A convergent hierarchy of semidefinite programs characterizing the set of quantum correlations},
	Url = {http://stacks.iop.org/1367-2630/10/i=7/a=073013},
	Volume = {10},
	Year = {2008}}

@article{ambainis2009quantumrandomaccesscodes,
      title={Quantum Random Access Codes with Shared Randomness}, 
      author={Andris Ambainis and Debbie Leung and Laura Mancinska and Maris Ozols},
      year={2009},
      eprint={0810.2937},
      archivePrefix={arXiv},
      primaryClass={quant-ph},
      doi={10.48550/arXiv.0810.2937}, 
}
\end{document}